\documentclass[useAMS,usenatbib]{mn2e}
\usepackage{latexsym,graphicx,natbib,amssymb,array,aas_macros}

\newcommand{\msun}{M_{\odot}}

\newcommand{\mpyr}{M_{\odot}\;{\rm yr}^{-1}}

\newcommand{\mscs}{M_{\rm SCS}}

\newcommand{\rhoecl}{\rho_{\rm ecl}}

\newcommand{\rh}{r_{\rm h}}

\newcommand{\fsing}{f_s}
\newcommand{\fb}{f_b}

\newcommand{\fbgal}{f_{b}^{\rm GF}}
\newcommand{\nb}{N_b}

\newcommand{\nbgal}{N_b^{\rm GF}}
\newcommand{\ns}{N_s}

\newcommand{\ncms}{N_{\rm cms}}

\newcommand{\ncmsgal}{N_{\rm cms}^{\rm GF}}
\newcommand{\nst}{N_*}
\newcommand{\lEb}{\log_{10}E_b}

\newcommand{\lP}{\log_{10}P}

\newcommand{\mecl}{M_{\rm ecl}}
\newcommand{\meclmin}{M_{\rm ecl,min}}
\newcommand{\meclmax}{M_{\rm ecl,max}}
\newcommand{\sfr}{{\rm SFR}}
\newcommand{\xiecl}{\xi_{\rm ecl}}
\newcommand{\kecl}{k_{\rm ecl}}
\newcommand{\dt}{\delta t}

\newcommand{\Eb}{E_b}

\newcommand{\mbar}{\overline{m}}

\newcommand{\Odyn}{\Omega_{\rm dyn}^{\mecl,\rh}}

\newcommand{\D}{{\cal D}^{\mecl,\rh}}

\newcommand{\Din}{{\cal D}_{\rm in}}

\newcommand{\Dgf}{{\cal D}_{\rm GF}}
\newcommand{\PhiE}{\Phi_{\lEb}^{\mecl,\rh}}
\newcommand{\PhiEin}{\Phi_{\lEb,\rm in}}

\newcommand{\tfreeze}{t_{\rm freeze}}

\title[Dynamical population synthesis]
      {Dynamical population synthesis: Constructing the stellar single and binary contents of galactic field populations}
      
\author[Michael Marks and Pavel Kroupa]
{
  Michael Marks$^{1,2,}$\thanks{Member of the International Max Planck Research School (IMPRS) for Astronomy and Astrophysics at the Universities of Bonn and Cologne; e-mail: mmarks@astro.uni-bonn.de (MM)} and Pavel Kroupa$^1$\\
  $^1$Argelander Institute for Astronomy, University of Bonn, Auf dem H\"ugel 71, 53121 Bonn, Germany\\
  $^2$Max-Planck-Institut f\"ur Radioastronomie, Auf dem H\"ugel 69, D-53121 Bonn, Germany\\
}       

\begin{document}

\date{Accepted ????. Received ?????; in original form ?????}

\pagerange{\pageref{firstpage}--\pageref{lastpage}} \pubyear{2010}

\maketitle

\label{firstpage}

\begin{abstract}
The galactic field's late-type stellar single and binary population is calculated on the observationally well-constrained supposition that all stars form as binaries with invariant properties in discrete star formation events. A recently developed tool (Marks, Kroupa \& Oh) is used to evolve the binary star distributions in star clusters for a few Myr until an equilibrium situation is achieved which has a particular mixture of single and binary stars. On cluster dissolution the population enters the galactic field with these characteristics. The different contributions of single stars and binaries from individual star clusters which are selected from a power-law embedded star cluster mass function are then added up. This gives rise to \emph{integrated galactic field binary distribution functions} (IGBDFs) resembling a galactic field's stellar content (Dynamical Population Synthesis). It is found that the binary proportion in the galactic field of a galaxy is larger the lower the minimum cluster mass, $\meclmin$, the lower the star formation rate, SFR, the steeper the embedded star cluster mass function (described by index $\beta$) and the larger the typical size of forming star clusters in the considered galaxy. In particular, period-, mass-ratio- and eccentricity IGBDFs for the Milky Way are modelled using $\meclmin=5\;\msun$, SFR$=3\;\mpyr$ and $\beta=2$ which are justified by observations. For $\rh\approx0.1-0.3$ pc, the half-mass radius of an embedded cluster, the afore mentioned theoretical IGBDFs agree with independently observed distributions, suggesting that the individual discrete star formation events in the MW generally formed compact star clusters. Of all late-type binaries, $50$~per~cent stem from $\mecl\lesssim300\msun$ clusters, while $50$~per~cent of all single stars were born in $\mecl\gtrsim10^4\msun$ clusters. Comparison of the G-dwarf and M-dwarf binary population indicates that the stars formed in mass-segregated clusters. In particular it is pointed out that although in the present model all M-dwarfs are born in binary systems, in the Milky Way's Galactic field the majority ends up being single stars. This work predicts that today's binary frequency in elliptical galaxies is lower than in spiral and in dwarf-galaxies. The period and mass-ratio distributions in these galaxies are explicitly predicted.
\end{abstract}


\section{Introduction}
\label{sec:intro}
The Milky-Way's (MW) Galactic field stellar late-type population (spectral-types G to M) consists of roughly $50$~per~cent single stars and $50$~per~cent binaries \citep{DuqMay1991,Mayor1992,FischerMarcy1992,Raghavan2010}. Throughout this work we will use  the \citet[hereafter DM91]{DuqMay1991} results for G-dwarfs as the canonical Galactic field population to which data is being compared, since it is the only existing long-term (13~yrs) survey performed by a single team, i.e. the DM91 data provides a homogeneous data set.

In a spectroscopic survey of halo, thick and thin disc populations, \citet{Carney2005} find for high proper motion stars that $28\pm3$~per~cent of all metal-poor ([Fe/H]$\leq-1$) centre-of-mass (cm-)systems\footnote{A \emph{system} refers either to a single star or to a binary.} are binaries. A similar number ($26\pm3$~per~cent) is found for the more metal-rich stars in their sample ([Fe/H]$>-1$). These values compare well with a binary-fraction of $22$~per~cent identified by DM91 over the same period range ($1.9<P<7500$~d). So there is no evidence for variation of binary properties in the Galactic field over cosmological time.

The origin of the Galactic field composition is a result of star formation but has not been predicted with success. Indeed, \citet{Fisher2004} finds in his theory of isolated star formation a bell-shaped binary period distribution function (period BDF) broadly similar to that of the Galactic field, but is not able to make specific predictions concerning the form of the period- and other BDFs. As discussed in \citet{Kroupa2011}, \citet{Moeckel2009}'s seminal SPH simulations result in a semi-major axis BDF with too many binaries with orbits around a few astromical units (AU). Furthermore, their mass-ratio distribution shows too few binaries with $q=m_2/m_1<0.8$, where $m_1$ and $m_2$ are the primary- and secondary-component mass, respectively. The reason for this might be a collapse which is too deep resulting in a too dense cluster and thus too efficient binary disruption. Additionaly after $10$~Myr of dynamical evolution including instantaneous residual-gas expulsion, orbits with semi-major axes $a>10$~AU are under-represented. Stellar feedback \citep[e.g.][]{Bate2009}, i.e. self-regulated star formation, might help to remedy this by reducing the depth of the collapse through an increased pressure.

Indeed, direct cloud collapse calculations are very limited in predicting binary star properties owing to the severe computational difficulties of treating the magneto-hydrodynamics together with correct radiation transfer and evolving atomic and molecular opacities during collapse. Such computations are also prohibitively expensive. There is therefore currently no sufficient numerical framework to derive the multiplicity properties of stars. While such computations are important for understanding the physical processes in observed star formation regions, they do not have predicitive power yet. In particular, larger systems, e.g. galaxies, cannot be synthesized.

Furthermore, the outcome of star formation computations cannot be expected to result in the binary properties of the Galactic field since clustered star formation is the dominant discrete star formation event \citep[e.g.][]{Lada2003,Lada2010,Bressert2010} and binaries are dynamically processed in star clusters before they become part of a galactic field. Therefore, a galactic field stellar population is the sum over all discrete star formation events, which on dissolution contribute a number of single stars and binaries dependend on the star formation conditions and dynamical history \citep{Kroupa1995b,Goodwin2010}. This is the topic of the present investigation.

Observations of low-density pre-main sequence populations, i.e. ``distributed`` star formation, show that most, if not all stars form as members of binaries and that they exhibit a period BDF that is rising towards long periods \citep{Simon1995,Ghez1997,Koehler1998,Duchene1999a,Connelley2008,Kroupa2011b}. Indeed, there is a simple but powerfull argument that the vast majority of stars must form in binaries: The lack of a significant single-star population in dynamically not evolved star forming regions means that stars cannot form in higher-order multiple systems. These would decay on a system crossing-time \citep[$10^5$~yr,][]{GoodwinKroupa2005}.

$N$-body computations of initially binary-dominated star clusters have shown that a rising period BDF can be turned into a bell-shaped one by gravitational interactions among the cluster members within a few Myr \citep[termed \emph{stimulated evolution},][Marks, Kroupa \& Oh 2011, Oh et al., in prep.]{Kroupa1995a,Kroupa1995b}. Here, an analytical treatment for the change of orbital-parameter BDFs in star clusters, which initially obeyed an invariant rising period BDF is used (Marks et al. 2011, hereafter Paper~I) to efficiently calculate the stellar single and binary content in individual star clusters. Then the populations coming from all star clusters of a galaxy's freshly formed star cluster system are summed which yields galactic field stellar populations once the star clusters have dissolved (\emph{Dynamical Population Synthesis}).

We note that the method developed here to calculate the integrated galactic-field binary distribution function (IGBDF, eq.~\ref{eq:igbdf} below) underlies similar concepts as the theory of the \emph{integrated galactic (stellar) initial mass function} (IGIMF), which sums up the IMFs in all discrete star formation events showing that galaxy-wide IMFs are steeper at the high-mass end than the invariant IMF in star clusters \citep{KroupaWeidner2003,WeidnerKroupa2005}. This theory has proven extraordinarily successfull in describing and predicting observational properties of galaxies \citep{Koeppen2007,Pflamm2008,Pflamm2009a,Recchi2009,Pflamm2009b,Calura2010}.

In Sec. \ref{sec:model} the model to calculate \emph{integrated galactic field binary distribution functions} (IGBDFs) is devised and in Sec. \ref{sec:results} the results are presented and compared to observations. Finally, Sec. \ref{sec:predict} discusses and shows model predictions and Sec. \ref{sec:sum} summarizes the main points of this investigation.

\section{Model}
\label{sec:model}
In order to integrate over all stellar populations in discrete star formation events (Sec.~\ref{sec:igbdf}), i.e. in embedded star clusters, it is necessary to first understand the evolution of binary populations in them (Sec.~\ref{sec:clusterbdf}). Note that the term discrete star formation event, star cluster and embedded star cluster are here used synonymously.

In the following reference will be made to \emph{birth or pre-main sequence} and \emph{initial} BDFs. The \emph{birth} distributions \citep{Kroupa1995a}, including random-pairing of component masses for late type stars, describe the properties of binary populations after they were born. But binaries are still embedded in their circum-stellar material, i.e. the components have not yet reached the main sequence stage. The \emph{initial} distributions \citep{Kroupa1995b} describe the corresponding statistical properties of a young binary population after birth binaries have undergone a phase of re-distribution of energy and angular momentum within their circumstellar material, called \emph{pre-main sequence eigenevolution}, acting on a time-scale of $<10^5$~yr. This mechanism introduces correlations between the orbital-parameters of short-period binaries \citep[such as between period and eccentricity,][]{Kroupa1995b} as seen in observations. For a summary of these processes the reader is referred to Paper~I.

\subsection{Binary distributions in star clusters}
\label{sec:clusterbdf}
In order to construct the field population by adding up the single stars and binaries in individual clusters, the evolution of binary populations has to be understood in terms of the initial properties of their host cluster $(\mecl,\rh)$.

In Paper~I an efficient method is provided to analytically describe the first $5$ Myr of the evolution of orbital-parameter BDFs in $N$-body computations of star clusters. The computations start with $100$~per~cent binaries distributed according to an initially rising period BDF derived in \citet{Kroupa1995b}, which is consistent with constraints for pre-main sequence and Class I protostellar binary populations (see Paper~I). Their method quantifies a \emph{stellar dynamical operator} \citep{Kroupa2002,Kroupa2008}, $\Odyn(t)$, in dependence of the initial cluster mass density, $\rhoecl\equiv3\mecl/8\pi\rh^3$, where $\mecl$ is the total mass in stars that formed in the embedded cluster and $\rh$ is its initial half-mass radius. This operator transforms an initial ($t=0$), perhaps invariant (see Paper~I), orbital-parameter BDF, $\Din$, into an evolved one, $\D(t)$, after some time $t$ of stimulated evolution,
\begin{equation}
 \D(t)=\Odyn(t)\times\Din\;.
 \label{eq:odyn}
\end{equation}
In the formulation of Paper~I $\Odyn$ acts in particular on the initial BDF for binding-energies, $\PhiEin$, i.e.
\begin{equation}
 \PhiE(t)=\Odyn(t)\times\PhiEin\;,
 \label{eq:odyn2}
\end{equation}
but extraction of other BDFs (period, semi-major axis, mass-ratio, eccentricity) is also possible with their model, given the interrelation between the orbital-parameters via Kepler's laws (Sec.~\ref{sec:extraction} below).

Let $\ncms=\ns+\nb$ be the number of systems, i.e. the sum of all single stars and binaries with primary-star mass near $m_1$ in a stellar population. Then a BDF, $\Phi^{\mecl,\rh}_x(m_1)\;(x=\log_{10}E,a,e,q,\ldots)$, is defined as the distribution of binary-fractions, $\fb=\nb/\ncms$, as a function of $x$,
\begin{equation}
 \Phi^{\mecl,\rh}_x(m_1)=\frac{d\fb(x)}{dx}=\frac{1}{\ncms}\frac{d\nb(x)}{dx}\;,
 \label{eq:bdf}
\end{equation}
where $d\nb(x)$ is the number of binaries in the interval $[x,x+dx]$. The total binary fraction equals the area below the BDF,
\begin{equation}
 \fb(m_1)=\int_{-\infty}^{\infty}\Phi^{\mecl,\rh}_x(m_1)\;dx\;.
 \label{eq:bdfnorm}
\end{equation}

\subsection{Integrated orbital-parameter distributions}
\label{sec:igbdf}
The galactic field's binary population is the sum over the populations in all star clusters that ever formed in a galaxy, having evolved for at least a time-span $\tfreeze$ after which the binary orbital-parameter properties become \emph{frozen-in},
\begin{equation}
\label{eq:dgfsimpleomega}
 \Dgf^{\rh}=\int_{\meclmin}^{\meclmax}\Odyn(\tfreeze)\;\Din\;\xiecl(\mecl)\;d\mecl\;,
\end{equation}
where $\xiecl(\mecl)$ is the embedded cluster mass function (ECMF). For simplicity it is assumed that all star clusters have formed with comparable half-mass radii, $\rh$. The integration ranges from a minimum-cluster-mass, $\meclmin$ to some maximum-cluster-mass, $\meclmax\equiv\meclmax(\sfr)$. The maximum-cluster-mass that can form in a galaxy which has a given star formation rate (SFR) is determined by \citep{Weidner2004},
\begin{equation}
 \label{eq:weidner}
 \frac{\meclmax(\sfr)}{\msun}=84793\times\left(\frac{\sfr}{\msun\;{\rm yr}^{-1}}\right)^{0.75}\;.
\end{equation}
Star clusters are distributed according to a power-law ECMF with index $\beta$,
\begin{equation}
 \xiecl(\mecl)=\kecl\times\mecl^{-\beta}\;,
 \label{eq:ecmf}
\end{equation}
which is normalized such that the sum of the masses of all clusters equals the total mass of the freshly formed star cluster system,
\begin{equation}
 \mscs=\int_{\meclmin}^{\meclmax(\sfr)}\mecl\;\xiecl(\mecl)\;d\mecl\;.
 \label{eq:norm}
\end{equation}
The total mass, $\mscs$, needed to find the normalization constant $\kecl$ is determined from the SFR and the formation time-scale, $\dt$, of the star cluster system,
\begin{equation}
 \mscs=\sfr\times\dt\;.
 \label{eq:mscs}
\end{equation}
\citet{Weidner2004} found that about every $\dt=10$ Myr of ongoing star formation an ECMF is fully populated.

The binary population will enter a galactic field with characteristics set at time $\tfreeze$. This is reached after sufficient time for cluster internal stimulated evolution when only \emph{hard binaries} are left (Paper~I), or if a cluster suddenly expands rapidly, e.g. as a result of residual-gas expulsion, which inhibits further stimulated evolution. This state is already reached after a few Myr or even in less than $1$~Myr for dense configurations \citep[Paper~I]{Kroupa1995a,Duchene1999b,Fregeau2009,Parker2009}. For our purpose, we choose the freeze-in time to coincide with the time of the occurence of the first supernovae, $\tfreeze\approx3$ Myr. If the cluster is still embedded, supernovae are expected to drive out the residual-gas of the embedded cluster rapidly, leading to cluster expansion and destruction of the majority of clusters in a star cluster system \citep[$\approx$90~per~cent, e.g.][]{Lada2003}. The exact time star clusters are allowed to evolve their population is not that important since the time-scale on which binary dissolution occurs is short. In particular, the difference in the binary fraction between $3$ and $5$~Myr in the $N$-body models used in Paper~I is of the order of a few per~cent only.

\subsection{Normalization}
\label{sec:norm}
Since the stellar-dynamical operator (eq.~\ref{eq:odyn2}) transforms between BDFs independently of the number of systems in a star cluster, one has to take care to preserve the definition and normalization for individual clusters (eqs.~\ref{eq:bdf} and~\ref{eq:bdfnorm}) also for the integrated population. Consider therefore as an example two clusters consisting of $\ncms=10$ and $100$ systems, with $\fb=40$ and $60$~per~cent, respectively. Evaluating eq.~(\ref{eq:dgfsimpleomega}) for the two clusters would result in $\fb=50$~per~cent for the combined (or integrated) population, but the true resulting population has $\fb=64/110=58$~per~cent. Thus, the number of systems making up the population has to be taken into account.

Define a BDF for a galaxy analogous to eq.~(\ref{eq:bdf}),
\begin{equation}
 \Phi_{x}^{\rm GF}(m_1)=\frac{d\fbgal(x)}{dx}=\frac{1}{\ncmsgal}\frac{d\nbgal(x)}{dx}\;,
 \label{eq:gfbdf}
\end{equation}
where $d\nbgal(x)$ and $\ncmsgal$ are the number of binaries in the interval of size $dx$ and the number of cm-systems in a whole galaxy, respectively. The goal is thus to calculate the number of binaries per $x$-interval and number of systems (single+binary) in a galaxy from the respective numbers in star clusters separately.

For an initial cluster mass, $\mecl$, the total number of freshly hatched stars in that particular cluster is calculated from
\begin{equation}
 \nst(\mecl)=\frac{\mecl}{\mbar}\;,
\end{equation}
where $\mbar\approx0.4\msun$ is the average mass of the canonical stellar IMF \citep{Kroupa2001}. The total number of binaries is related to $\nst(\mecl)$ via
\begin{equation}
 \nb(\mecl)=\nst(\mecl)-\ncms(\mecl)\;.
 \label{eq:nbtot}
\end{equation}
Inserting this in $\fb=\nb/\ncms$ and rearranging gives the number of cm-systems in that particular cluster,
\begin{equation}
 \ncms(\mecl)=\frac{\nst(\mecl)}{1+\fb}\;.
 \label{eq:ncms}
\end{equation}

Since $\Phi^{\mecl,\rh}_x$ is known (eq.~\ref{eq:odyn2}) and therefore $\fb$ (eq.~\ref{eq:bdfnorm}), $\ncms$ and $\nb$ can be calculated. The number of binaries per interval $dx$ becomes,
\begin{equation}
 \frac{d\nb(\mecl)}{dx}=\ncms(\mecl)\times\Phi^{\mecl,\rh}_x\;.
 \label{eq:nb}
\end{equation}

Therefore, eq.~(\ref{eq:gfbdf}) becomes,
\begin{equation}
 \Phi_{x}^{\rm GF}(m_1)=\frac{\int_{\meclmin}^{\meclmax(\sfr)}\frac{d\nb(\mecl)}{dx}\;\xiecl(\mecl)\;d\mecl}{\int_{\meclmin}^{\meclmax(\sfr)}\ncms(\mecl)\;\xiecl(\mecl)\;d\mecl}\;.
 \label{eq:igbdf}
\end{equation}
Eq.~(\ref{eq:igbdf}) is referred to as the \emph{integrated galactic-field binary distribution function} (IGBDF). It can be calculated for single stars and binaries with primary-mass in an interval $\Delta m$ around $m_1$, e.g. $0.8-1.04\msun$, or for all late-type stellar systems $m_1/\msun\in[0.08,2]$ (Tab.~\ref{tab:spectypes}).

\subsection{P-, e-, q- and a-IGBDF}
\label{sec:extraction}
In order to extract IGBDFs for different orbital-parameters (period, $P$, eccentricity, $e$, mass-ratio, $q$, and semi-major axis, $a$) from the known energy IGBDF, the procedure is basically as described in sec. 4.2 of Paper~I for orbital-parameter distributions in single clusters. The idea is to compile a large library of binaries ($N_{\rm lib}=10^7$ binaries for the present purpose) whose properties are selected according to the recipe in \citet[see Paper~I]{Kroupa1995b}. Here, the library consists only of binaries with primary masses up to $2\msun$ in order to resemble a galactic field and it contains the values for $m_1,m_2,\Eb,P,e,q$ and $a$. The total number of initial binaries in the integrated population, calculated according to Sec. \ref{sec:norm}, is scaled so as to match the library size of $N_{\rm lib}$ binaries. Following this, the final distributions are constructed by removing an appropriate amount of binaries with a given binding-energy from the library according to the number-ratio of binaries in the resulting and initial energy IGBDFs. The remaining binaries are used to construct the $P-,e-,q-$ and $a-$IGBDF. In order to extract sub-distributions, such as for a special spectral-type or period-range, only those remaining binaries are used which fullfill the additional criteria. In the forthcoming sections, mass-ranges for single-stars or primary-component masses for binary-stars, respectively, are adopted, as shown in Tab.~\ref{tab:spectypes}.
\begin{table}
\begin{center}
\caption{Adopted mass-ranges for single-stars or primary-components of binaries with different spectral-type (SpT).}
\begin{tabular}{r||c|c|c|c}
 SpT & F & G & K & M \\
\hline
 $m_*/\msun$ & $1.04$-$1.4$ & $0.8$-$1.04$ & $0.45$-$0.8$ & $0.08$-$0.45$
\end{tabular}
\label{tab:spectypes}
\end{center}
\end{table}

\section{Results}
\label{sec:results}
According to eq.~(\ref{eq:igbdf}), the properties of an IGBDF depend on the minimum embedded cluster mass and the SFR (i.e. the maximum cluster mass, eq.~\ref{eq:weidner}) of the considered galaxy, the index, $\beta$, of the ECMF (eq.~\ref{eq:ecmf}), and the average half-mass radius, $\rh$, of the star clusters. The influence of these parameters on the integrated binary properties by means of the global binary fraction will be investigated. A discussion of the Galactic field binary population of the Milky-Way (MW) and what can be learned about the initial star cluster system of the MW from the observed binary population ends this section.

\subsection{The importance of low-mass clusters for the Galactic field binary population}
\label{sec:importance}
\begin{figure*}
 \begin{center}
 $\begin{array}{cc}
   \includegraphics[width=0.45\textwidth]{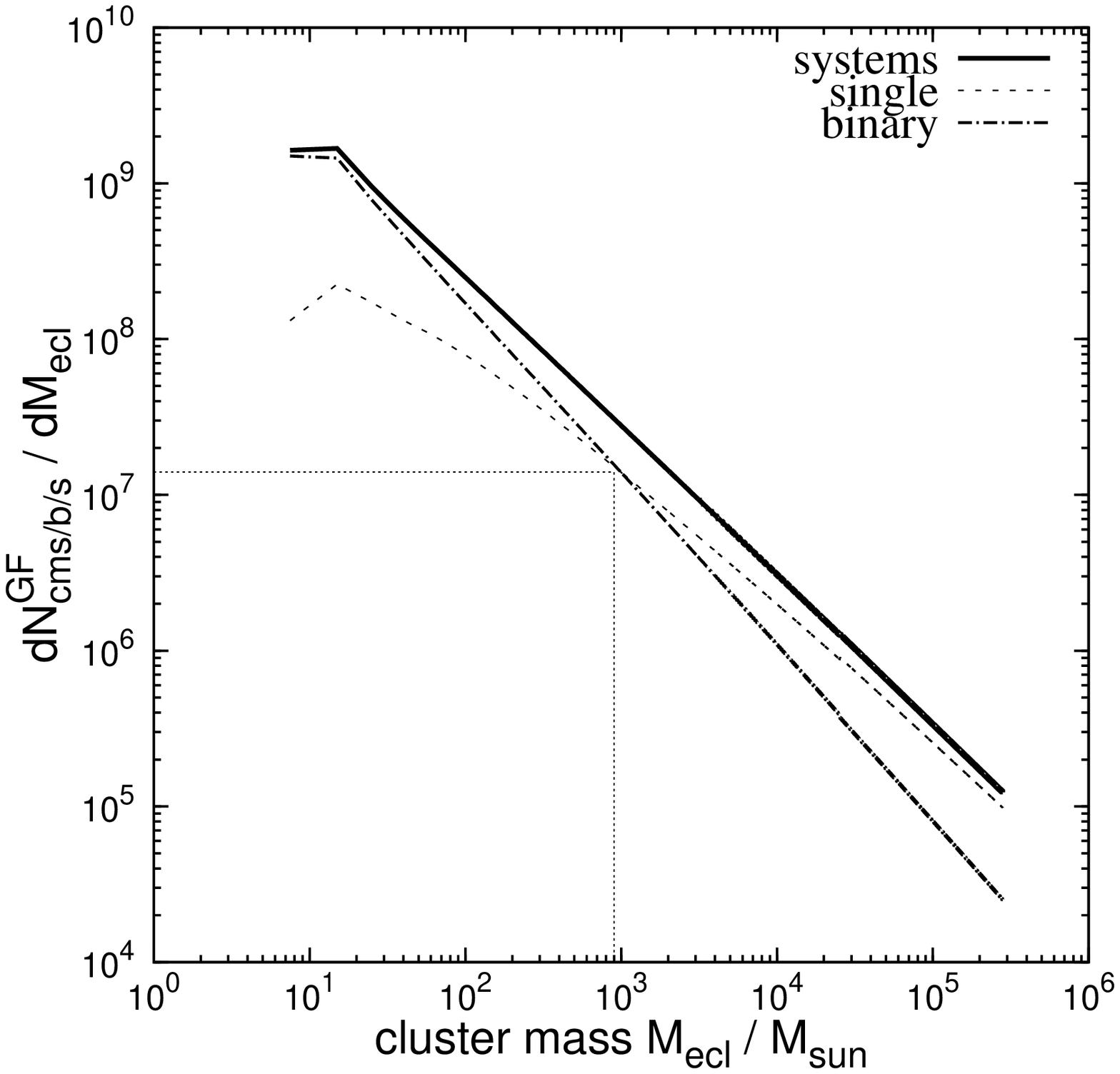} &
   \includegraphics[width=0.45\textwidth]{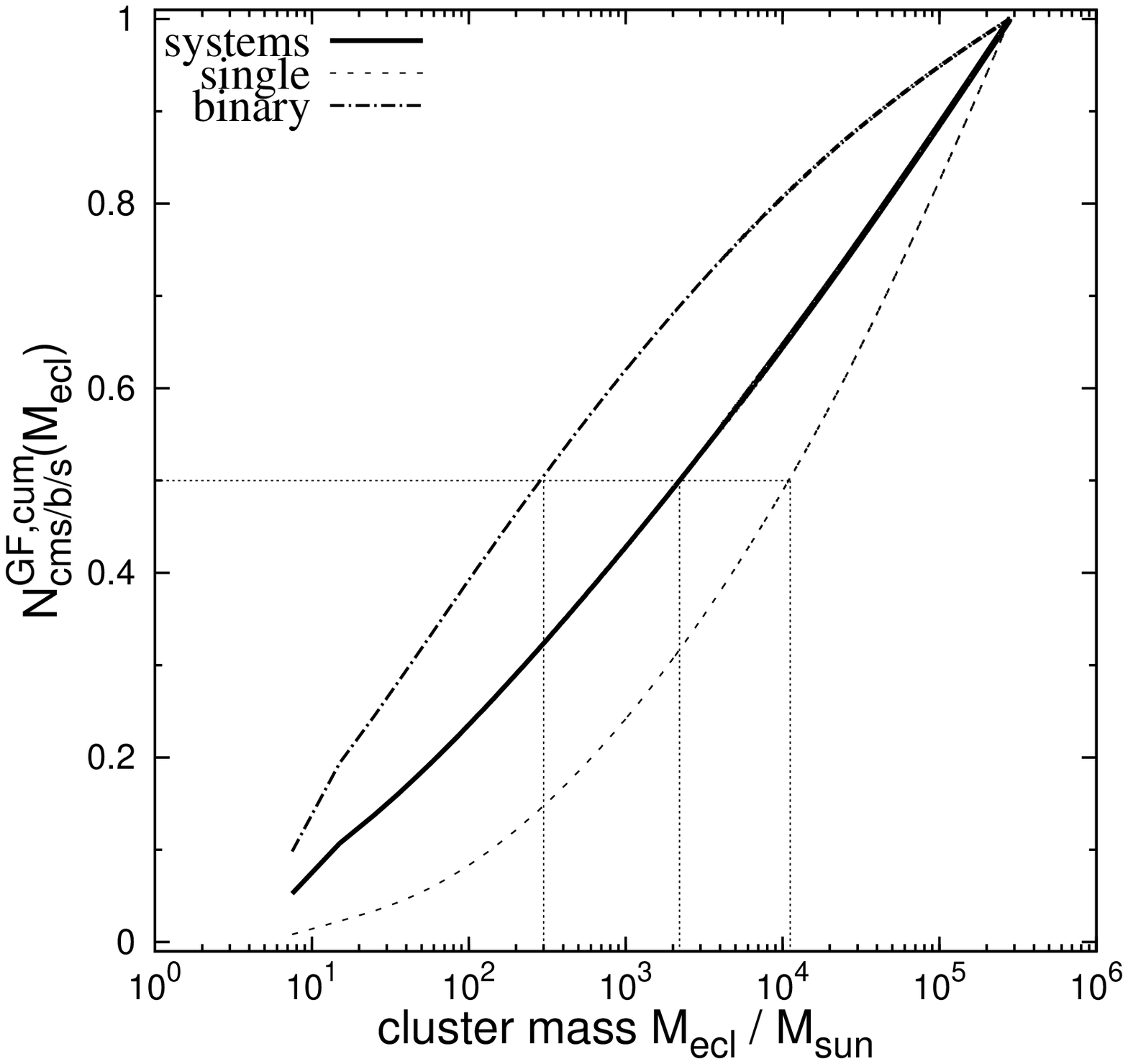}
 \end{array}$
 \end{center}
  \caption{\textbf{Left~panel:} Number of all singles (dotted), binaries (dash-dotted) and all systems (single+binary, solid) per cluster mass in the Galactic field (eq.~\ref{eq:nmecl}), respectively, that form within one star cluster system formation time-scale $\delta t$ (eq.~\ref{eq:mscs}). Lines are drawn for the MW star cluster system ($\beta=2.0$, $\rh=0.2$~pc, $\meclmin=5\msun$ and SFR=$3\mpyr$). Low-mass clusters contribute most systems to the Galactic field population owing to the steepness of the ECMF. The intersection between the single-stars and binary-stars line is the cluster mass at which $\fb=50$~per~cent for the used parameters. \textbf{Right~panel:} Cumulative number of singles, binaries and all systems as a function of $\mecl$ normalized to the respective total number (eq.~\ref{eq:ncum}) for the same star cluster system parameters as in the left panel. Low-mass clusters ($\mecl\lesssim300\msun$) are the dominant binary contributors while high-mass clusters ($\mecl\gtrsim10^4\msun$) donate most single stars.}
 \label{fig:importance}
\end{figure*}
From which type of cluster do most Galactic field binaries originate? Dissolving low-mass embedded clusters will each contribute only a small number of systems to a Galactic field population but a large fraction of binaries will be among them due to inefficient stimulated evolution (Paper~I). For high-mass clusters the situation is the other way round. However, the ECMF (eq.~\ref{eq:ecmf}) is steep, i.e. there are many more low-mass than high-mass clusters. Thus, this question is not simple to answer qualitatively.

Therefore the number of systems, binaries and single stars which are added to the Galactic field by all clusters of a given mass $\mecl$ is calculated by evaluating
\begin{equation}
 \frac{dN_{{\rm cms}/b/s}^{\rm GF}}{d\mecl}=N_{{\rm cms}/b/s}(\mecl)\;\xiecl(\mecl)\;.
 \label{eq:nmecl}
\end{equation}
The result is depicted in the left panel of Fig.~\ref{fig:importance} for the MW star cluster system with $\beta=2.0$, $\rh=0.2$ pc, $\meclmin=5\msun$ and SFR=$3\mpyr$ (see Sec.~\ref{sec:MW} below). It shows that the steepness of the ECMF dominates such that most systems in the Galactic field stem from low-mass clusters. For the same star cluster system parameters, the right panel of Fig.~\ref{fig:importance} plots the normalized cumulative number,
\begin{equation}
 N^{\rm GF,cum}_{{\rm cms}/b/s}(\mecl)=\frac{1}{N_{{\rm cms}/b/s}^{\rm GF}}\int_{\meclmin}^{\mecl}\frac{dN_{{\rm cms}/b/s}^{\rm GF}}{d\mecl}d\mecl
 \label{eq:ncum}
\end{equation}
where $N_{{\rm cms}/b/s}^{\rm GF}$ is the total number of systems, binaries or singles in the Galactic field, respectively, i.e. eq.~(\ref{eq:nmecl}) integrated from $\meclmin$ to $\meclmax$. It demonstrates that about $50$~per~cent of all systems in the Galactic field formed in clusters with initial masses $\mecl\lesssim2\times10^3\msun$, while $50$~per~cent of all Galactic field binaries originate from clusters with $\mecl\lesssim300\msun$ only. For single stars the situation is inverted. Roughly $50$~per~cent of all single stars come from star clusters with masses $\mecl\gtrsim10^4\msun$.

\subsection{Parameter study \& degeneracy}
\label{sec:paramstudy}
\begin{figure*}
 \begin{center}
 $\begin{array}{ccc}
   \includegraphics[width=0.3\textwidth]{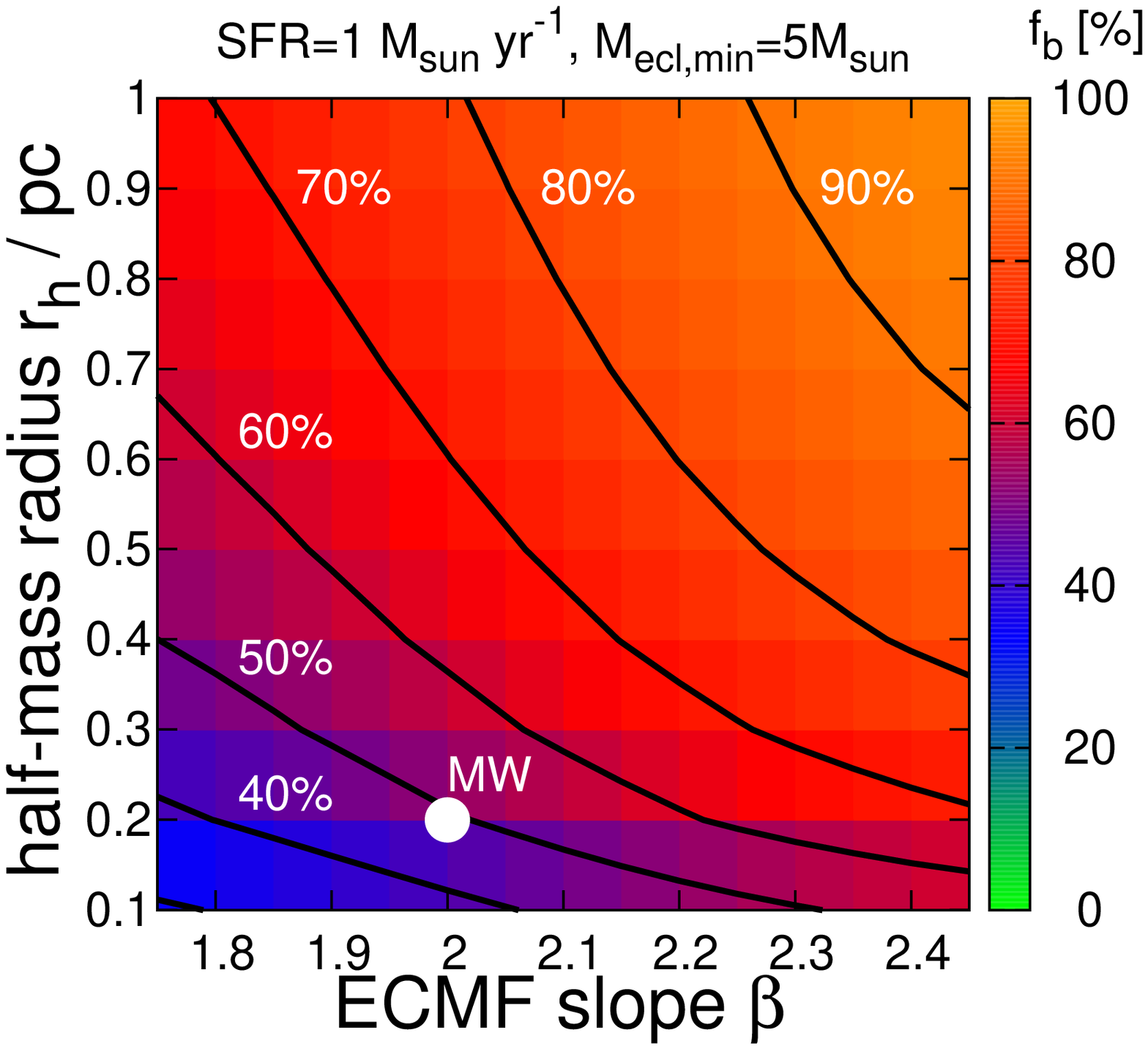} & \includegraphics[width=0.3\textwidth]{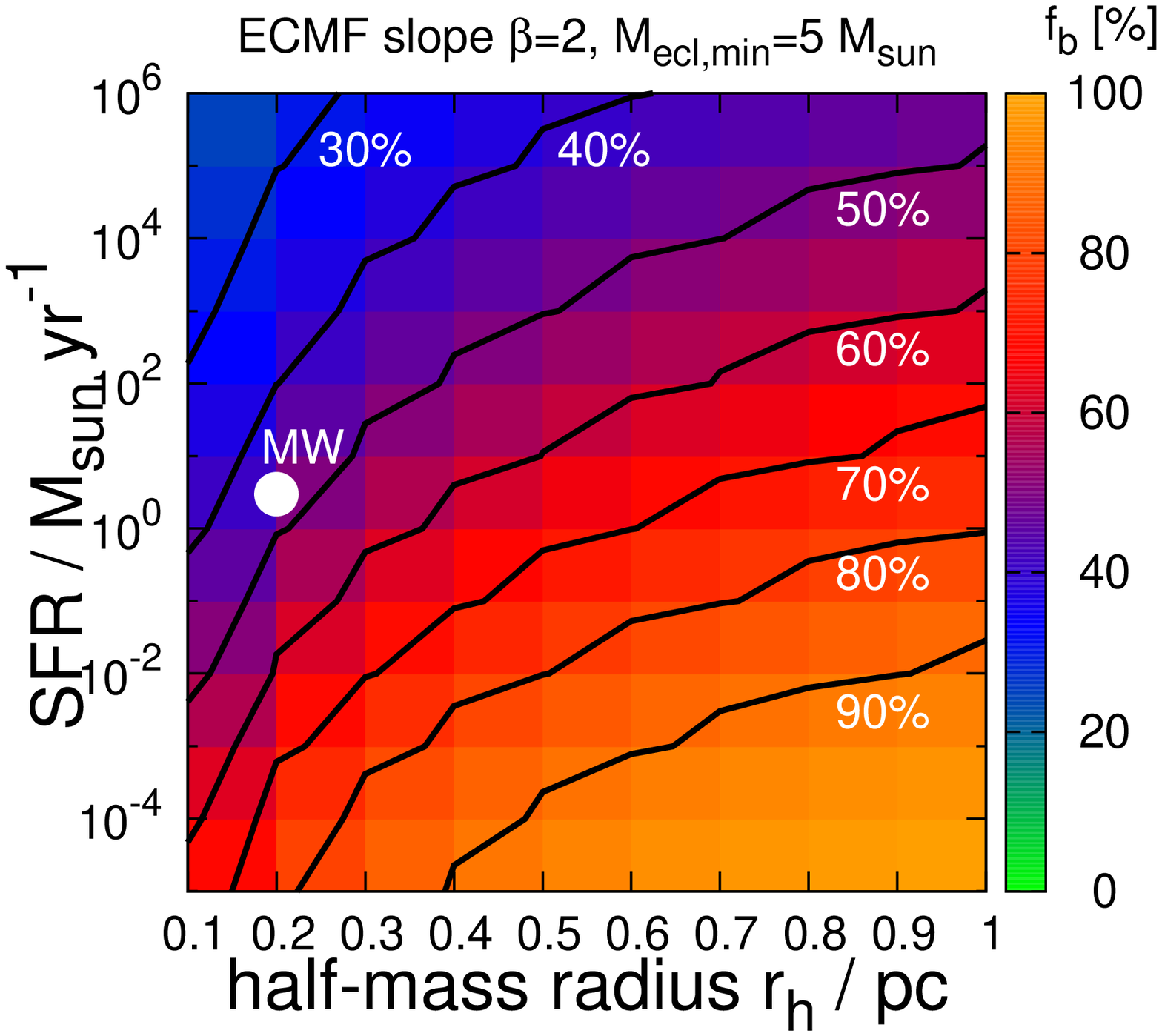} & \includegraphics[width=0.3\textwidth]{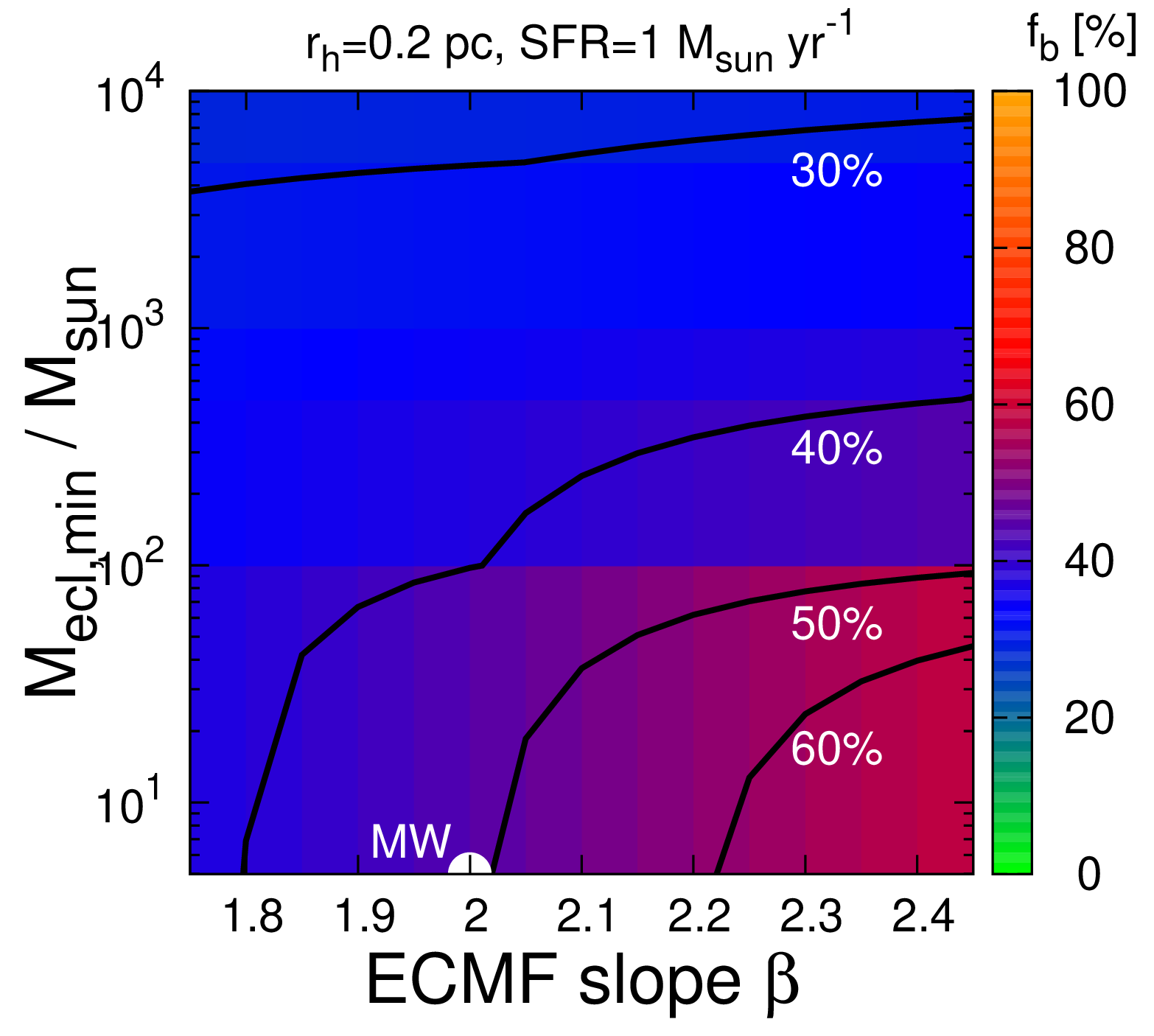}
 \end{array}$
 \end{center}
  \caption{Study of the influence of the parameters in eq. (\ref{eq:igbdf}) determining the integrated stellar population. In each panel the color-coding shows the global binary-fraction, $\fb$, i.e. including all late-type binaries ($m_1\leq2\msun$), according to the bar on the right of each panel (in per~cent). The solid overlaid lines indicate curves of constant binary fraction. The big solid white dot in each panel marks the position of the MW (Sec.~\ref{sec:MW}). The binary population increases with decreasing SFR, with increasing cluster half-mass radii, $\rh$, with steepening of the ECMF (increasing index $\beta$) and with decreasing minimum cluster mass, $\meclmin$.}
 \label{fig:paramstudy}
\end{figure*}
Fig.~\ref{fig:paramstudy} depicts the global binary fraction (color-coded), i.e. including all late-type binaries ($m_1\leq2\msun$), of a galaxies' stellar population in dependence of the four parameters $(\meclmin,{\rm SFR},\beta,\rh)$ in eq. (\ref{eq:igbdf}). We learn the following from these diagrams: If all other parameters are fixed...
\begin{itemize}
 \item[...] \textit{the higher the SFR, the smaller the field-binary population}. A higher SFR yields a larger $\mecl$ (eq. \ref{eq:weidner}). A higher-mass cluster is generally denser than a lower-mass cluster and will therefore contribute a population with a lower binary fraction (Paper~I). Therefore $\fb$ drops with increasing SFR.
 \item[...] \textit{the larger $\rh$, the larger the galactic field binary fraction}. Larger typical half-mass radii imply lower densities, resulting in dynamically less evolved binary populations before dissolution (higher $\fb$, Paper~I). Therefore $\fb$ increases with increasing $\rh$.
 \item[...] \textit{the steeper the ECMF (larger index $\beta$), the larger the field-binary population}. A steeper ECMF amounts to a lesser proportion of high-mass, i.e. higher-density clusters. Therefore on average populations with larger binary-fractions enter the field and $\fb$ increases.
 \item[...] \textit{the larger the minimum cluster mass, the fewer binaries exist in the field}. Increasing $\meclmin$ means cutting out of low-density clusters which would contribute populations with large binary proportions. The remaining higher-mass clusters shed dynamically more evolved populations into a galactic field.
\end{itemize}
Fig.~\ref{fig:paramstudy} distincly shows that the parameter space is degenerate. E.g., the increase in $\fb$ due to a steeper ECMF can be compensated for by introducing smaller cluster sizes (left panel) or assuming a larger $\meclmin$ (right panel). Similarly, a larger typical $\rh$ can be counteracted with a higher SFR (middle panel), and so on. Therefore constraints on at least some of the four parameters from independent sources are needed to reduce the allowed solution space, when comparing model predictions with observed distributions. Those are available for the MW.

\subsection{The Milky-Way}
\label{sec:MW}
In order to calculate a synthetic stellar population for the MW's Galactic field, we need to estimate the parameters entering the IGBDF for our Galaxy.

The current global SFR of the MW using different methods is determined to lie between $\approx1$ and $5\mpyr$ \citep{Smith1978,Diehl2006,Misiriotis2006,Calzetti2009,Murray2010,Robitaille2010}. These values are compatible with the assumption that the total stellar mass in the disk and bulge of the MW \citep[$5\times10^{10}\msun$,][]{BinneyTremaine2008} has assembled continously during the last $13.7$~Gyr with a SFR of $3.6\mpyr$. Assuming SFR=$3\mpyr$ yields $\meclmax=1.9\times10^5\msun$ (eq. \ref{eq:weidner}) for the most massive cluster having formed in the MW disk, comparable to the most massive open cluster in \citet{Piskunov2007}'s sample of 236 open clusters within $1$~kpc from the Sun ($1.1\times10^5\msun$ for Sco~OB5). Taking the minimum cluster mass similar to that of a Taurus-Auriga like group \citep[$\meclmin=5\msun$]{KroupaBouvier2003} and an ECMF index $\beta=2$ similar to the observed slopes for Galactic embedded clusters \citep{Lada2003,delaFuenteMarcos2004,Gieles2006}, we can evaluate the properties of the Galactic field binary population by numerically integrating eq.~(\ref{eq:igbdf}) for different embedded cluster half-mass radii.

\subsubsection{Energy distribution}
\begin{figure}
 \begin{center}
    \includegraphics[width=0.45\textwidth]{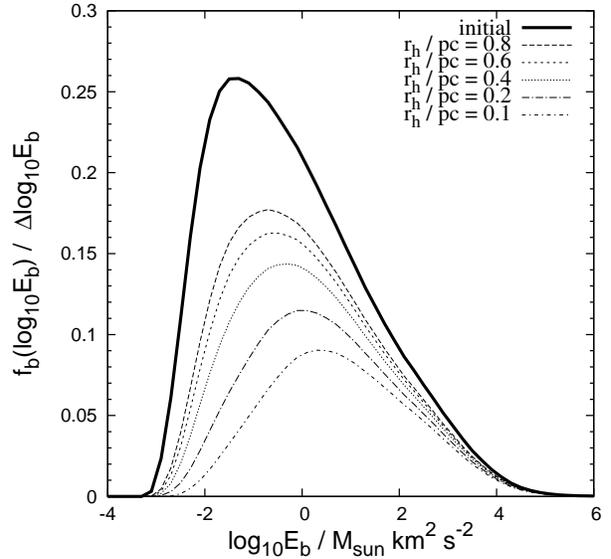}
 \end{center}
  \caption{Global energy IGBDFs for different typical initial cluster half-mass radii in the MW. The smaller the clusters are, the larger are their densities and the more efficient is the binary destruction initially (Sec. \ref{sec:paramstudy}). Therefore the binary fraction is smaller the lower the initial value of $\rh$ is, ranging from $\fb=0.34$ for $\rh=0.1$ pc to $\fb=0.71$ for $\rh=0.8$ pc. The area under each distribution equals the total model binary-fraction in the Galactic field.}
 \label{fig:mwlE}
\end{figure}
Fig.~\ref{fig:mwlE} shows the energy IGBDF for the above parameters, for different cluster half-mass radii and for primaries of all masses. It is found that the binary-fraction in the Galactic field (the area under the distributions) is smaller if the typical cluster radii are smaller. Smaller average radii lead to higher initial densities in star clusters which results in more efficient binary dissolution and thus a lower binary-fraction (Sec.~\ref{sec:paramstudy}). For $\rh=0.1$~pc the overall binary-fraction is $\fb=0.34$ and for $\rh=0.8$~pc, $\fb=0.71$.

\subsubsection{Period distribution}
\label{sec:mwlp}
\begin{figure*}
  \begin{center}
   $\begin{array}{cc}
     \includegraphics[width=0.45\textwidth]{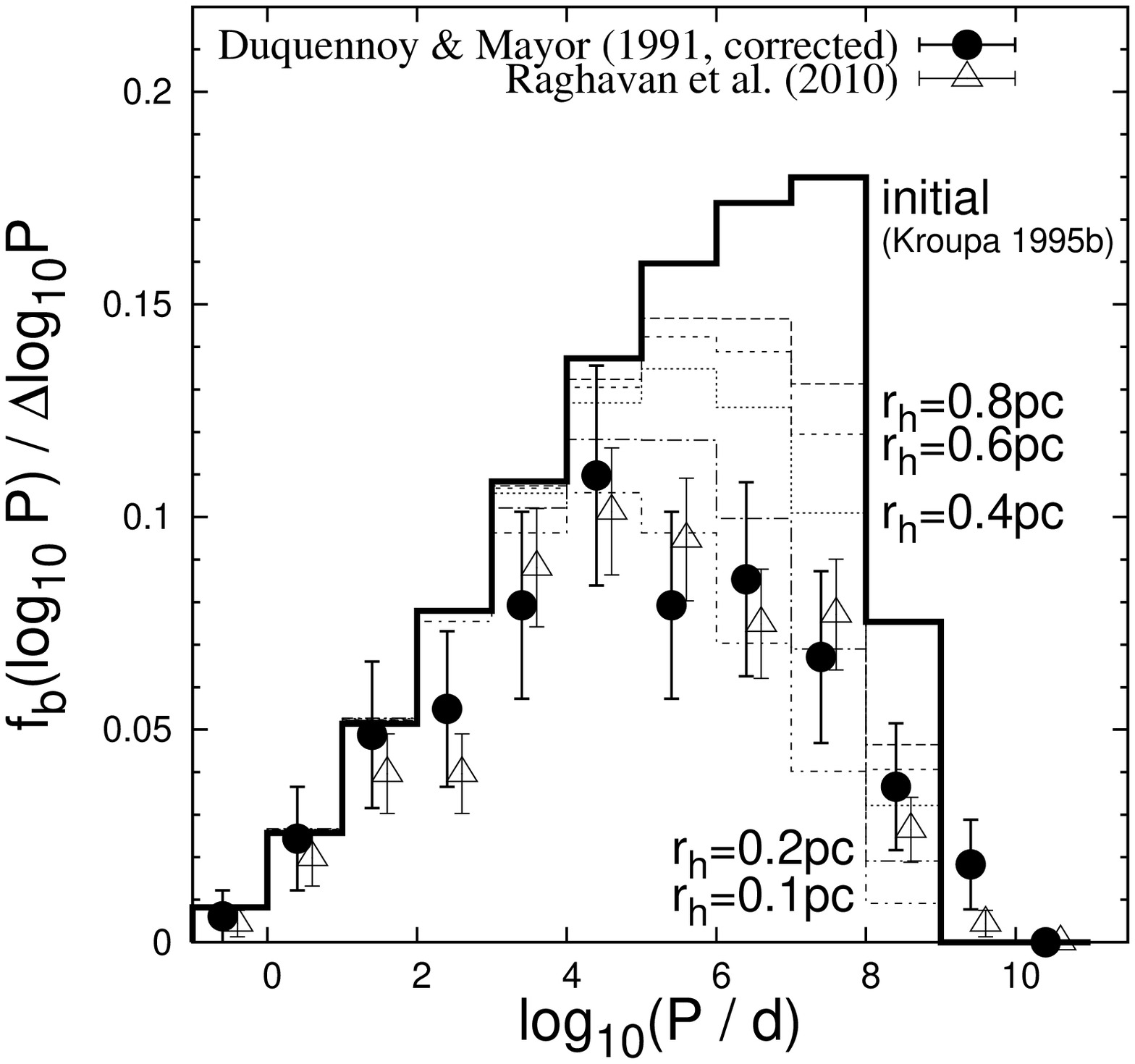} & \includegraphics[width=0.45\textwidth]{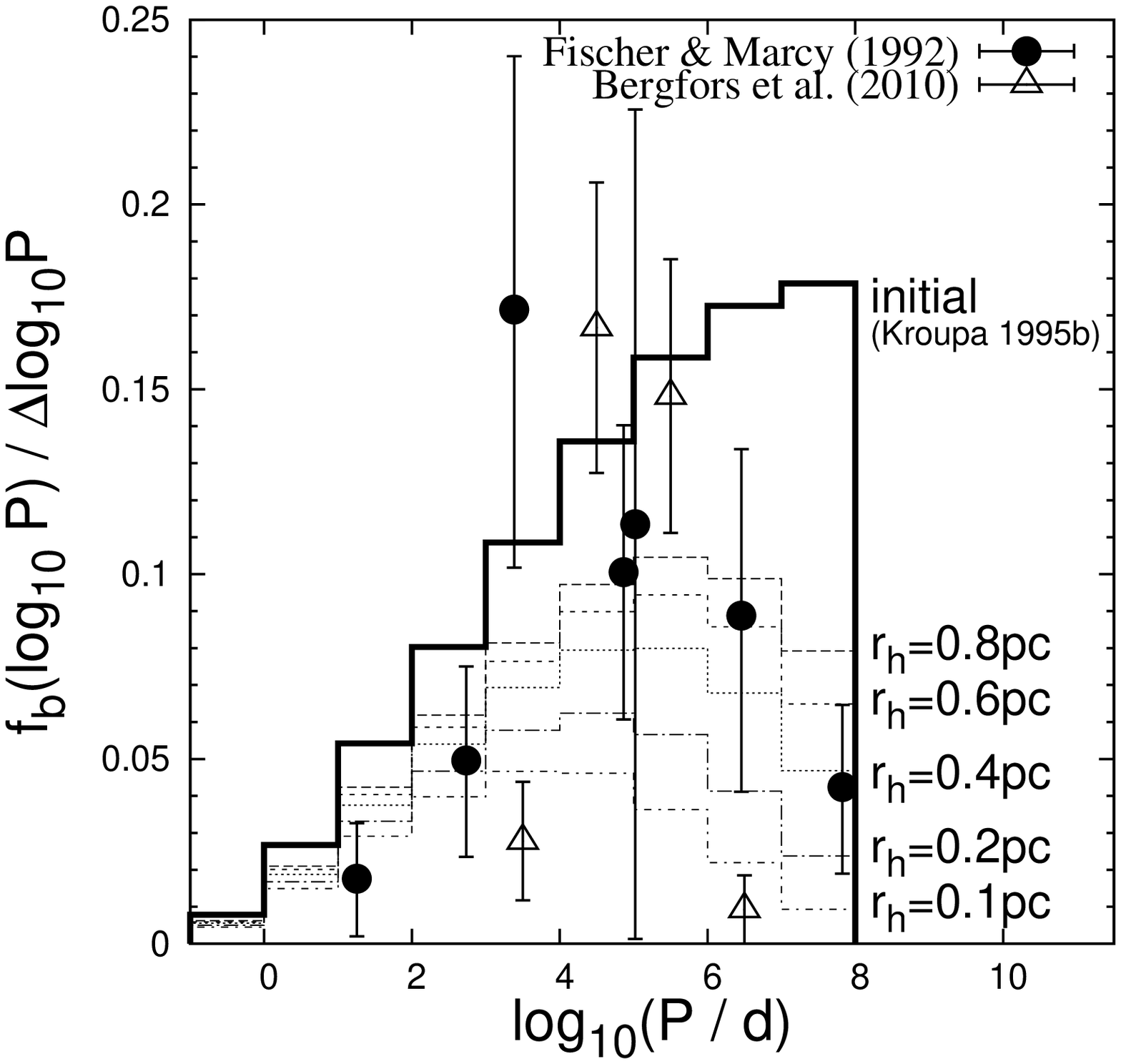}
   \end{array}$ 
  \end{center}
  \caption{Period IGBDFs for the MW model (Sec \ref{sec:MW}) with different typical initial cluster radii for G-dwarf (left) and M-dwarf (right) binaries (histograms). Comparison with the G-dwarf period BDF by DM91 (circles) and \citet[triangles]{Raghavan2010} suggests that clusters should have formed very compact. The period IGBDF with a half-mass radius of $\rh=0.1$~pc matches the observed binary-frequency best. The observed M-dwarf binary-frequency (\citet[robust]{FischerMarcy1992} and \citet{Bergfors2010}) favours formation in $\rh=0.3$~pc sized clusters. This difference in $\rh$ between the G-dwarf and M-dwarf solution is discussed in Sec.~\ref{sec:predict}. The line-types are as in Fig.~\ref{fig:mwlE}.}
 \label{fig:mwlP}
\end{figure*}
For the adopted initial conditions in star clusters (Paper~I) and the above values for the SFR, $\beta$ and $\meclmin$ in the MW, the period IGBDFs in Fig.~\ref{fig:mwlP} for different values of $\rh$ demonstrate that the initially rising period distribution in star clusters translates into a bell-shaped form in the Galactic field, at least if the star clusters from which the Galactic field population originates, were rather compact. Comparison with the corrected observed Galactic field period BDF for G-dwarfs (DM91) suggests that clusters formed quite compact (left panel). The G-dwarf field binary fraction of $\approx57$~per~cent is best reproduced with a typical initial cluster size of $\rh=0.1$~pc, where $\fb({\rm G})=0.58$. A similar study by \citet{Raghavan2010} investigating the multiplicity properties of $454$ solar-type stars selected from the Hipparcos catalogue is in agreement with the DM91 data and thus with the compact formation of star clusters.

\citet{FischerMarcy1992} studied the M-dwarf binary population within $20$~pc from the Sun, showing that the period distribution of M-dwarfs can also be described by a log-normal distribution. Comparison with the IGBDF models (Fig.~\ref{fig:mwlP}, right panel) favors their formation in slightly more extended clusters. For $\rh=0.3$~pc the binary-frequency of $41$~per~cent compares well with the observational result ($\fb=0.42$). Recently, \citet{Bergfors2010} surveyed 124 M-type stars for binary separations $a\lesssim200{\rm AU}$ for their multiplicity properties. The observed semi-major axis distribution has been translated into periods using Kepler's laws and are incorporated in Fig.~\ref{fig:mwlP} (right panel). Their data are compatible with the \citet{FischerMarcy1992} data. For a sub-sample of the model binary population with $a\lesssim200$~AU, as in the observations, the $\rh=0.2$~pc-model binary-frequency is $\fb=0.31$ and agrees with the observed binary-fraction of $\approx32$~per~cent. \citet{Bergfors2010} note, however, that there might be some overabundance of systems with $P<20$~d, corresponding to $a\lesssim80$~AU for an average system mass of $\approx0.5\msun$, due to the sample selection and that corrections might be necessary for non-physical (optical) pairs.

\subsubsection{Mass-ratio distribution}
\begin{figure*}
 \begin{center}
 $\begin{array}{ccc}
   \includegraphics[width=0.3\textwidth]{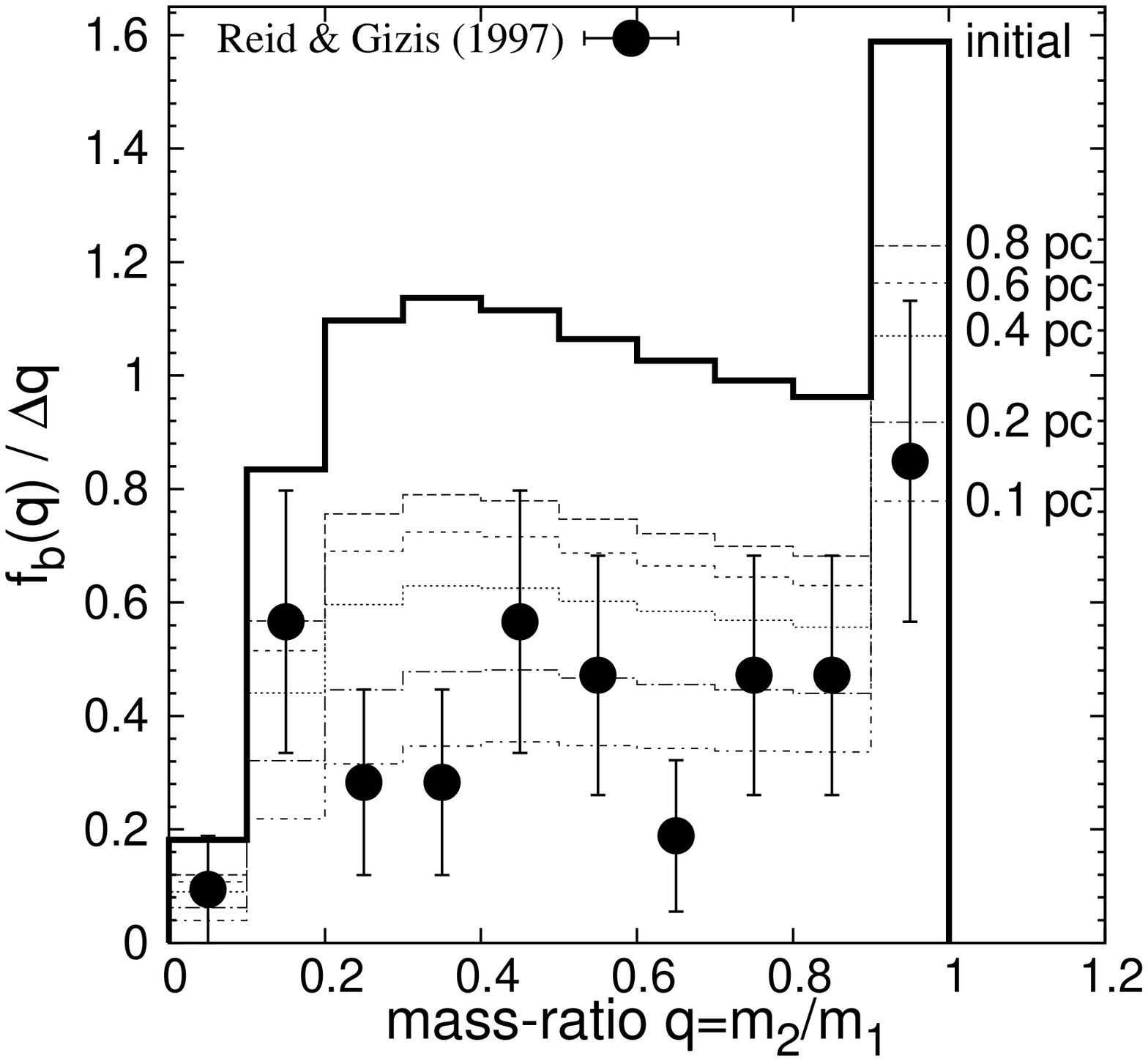} &
   \includegraphics[width=0.3\textwidth]{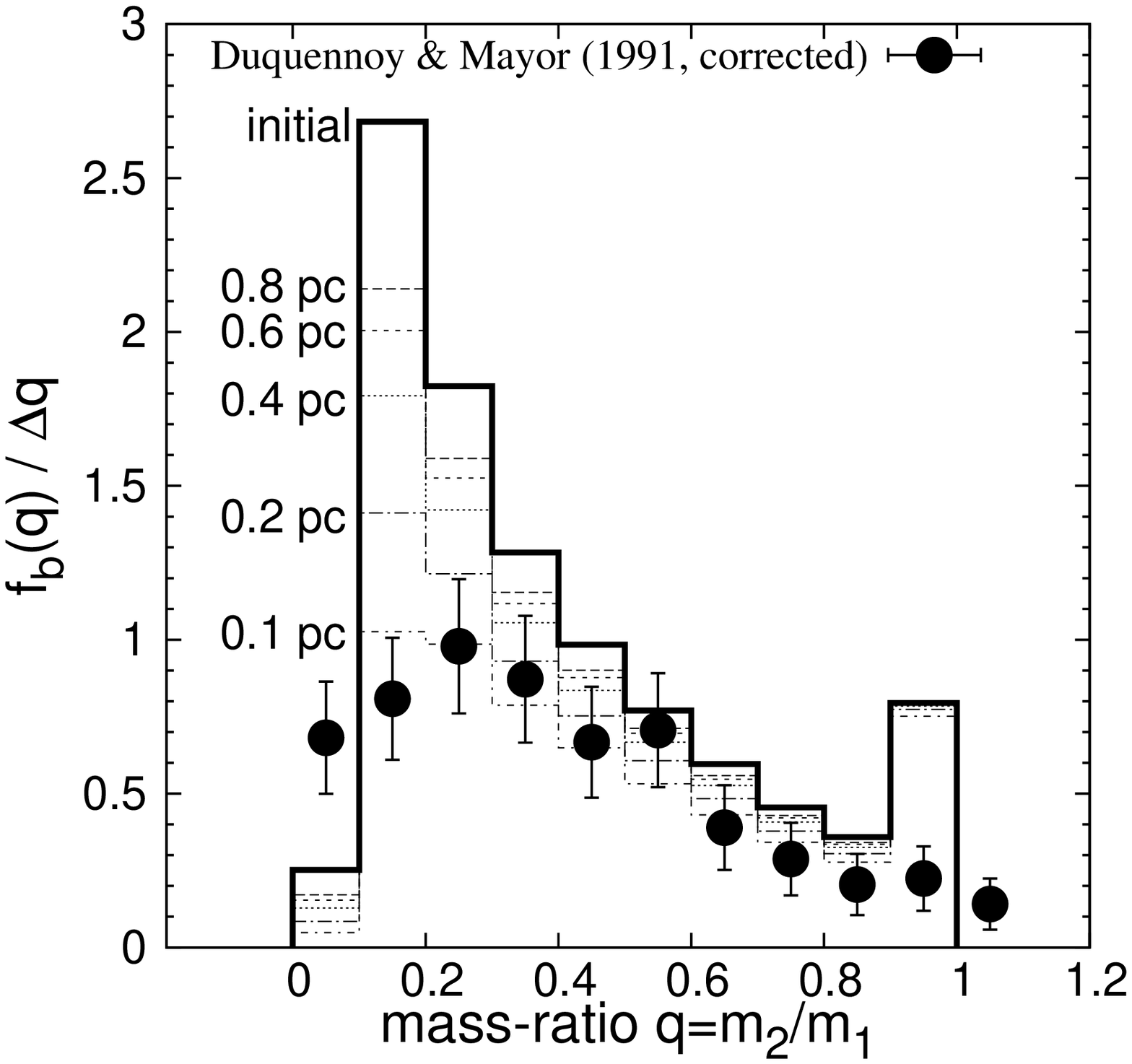} &
   \includegraphics[width=0.3\textwidth]{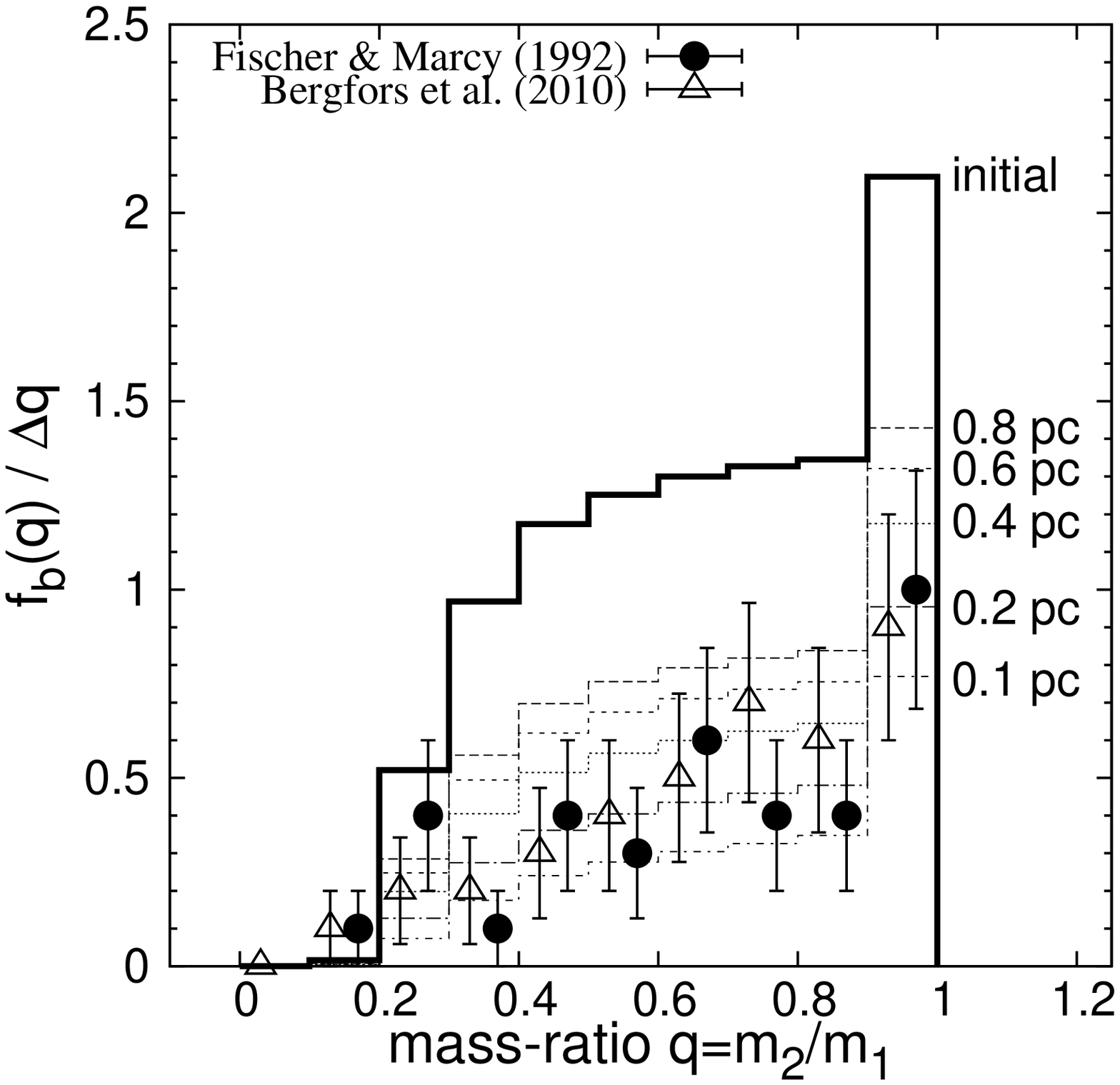} 
 \end{array}$
 \end{center}
 \caption{Mass-ratio IGBDFs for all late-type binaries ($m_1\leq2\msun$, left panel), G-dwarf (middle panel) and M-dwarf binaries (right panel), for the MW model (Sec. \ref{sec:MW}) with different half-mass radii (histograms). The line-types are as in Fig.~\ref{fig:mwlE}. While the distribution between $q=0.2$ and $0.9$ is flat for the complete distribution it declines for the G-dwarfs and increases for M-type binaries with increasing~$q$. The peak at mass ratio's in the model close to unity (an effect of pre-main sequence eigenevolution) is also evident in the observational data for all primary-masses combined \citep[left]{Reid1997} and M-dwarf binaries \citep[right]{FischerMarcy1992,Bergfors2010}, while it is less pronounced, if at all, for the G-dwarf observations by DM91 (middle). Note that agreement is obtained for the similar $\rh$ as for the period IGBDFs (Fig.~\ref{fig:mwlP}).}
 \label{fig:mwmr}
\end{figure*}
The result that star formation in rather compact structures is favoured is independently confirmed by considering the distribution of mass-ratios. Fig.~\ref{fig:mwmr} depicts the mass ratio IGBDFs for the complete field population (left, IGBDF for all binaries with primary masses $m_1\leq2\msun$), for G-dwarf binaries only (middle) and for M-dwarfs (right). It is apparent that the complete distribution is flat between roughly $q=0.2$ and $0.9$ while the G-type binary sub-distribution is decreasing and the M-dwarf IGBDF is increasing with increasing~$q$. We emphasize, that \emph{all three} mass ratio IGBDFs result from initially sampling the two \emph{birth} components of a binary randomly from the same underlying stellar IMF \citep[e.g.][]{Kroupa1995a}. The decreasing trend is also evident for F- and K-type primaries (Fig.~\ref{fig:predict} below) and is seen in the observations of G-dwarfs by DM91 (middle panel of Fig.~\ref{fig:mwmr}). The observational data compare very well with the compact formation models, as above for the period IGBDFs. Addition of the mass ratio IGBDFs for the different spectral types (Tab.~\ref{tab:spectypes} and Fig.~\ref{fig:predict} below) results in the flat distribution for the complete mass ratio IGBDF (left panel in Fig.~\ref{fig:mwmr}). The binary fraction of $35$~per~cent in the observations of $106$ G- to M-type systems in the analysis by \citet{Reid1997} are best compatible with the $\rh=0.1$~pc model where $\fb=0.34$.

In contrast to the declining $q$-distribution for G-dwarfs found by DM91, note that the results of \citet{Raghavan2010} suggest a mass-ratio distribution for binaries with a solar-type primary which is flat between $q\approx0.2$ and $0.9$, while finding a similar period distribution (see discussion in Sec.~\ref{sec:predict}). \citet{FischerMarcy1992} and \citet{Bergfors2010} extracted also a mass ratio distribution for their respective samples of M-dwarfs whose rising shapes are well reproduced by the models (right panel). As expected from the period IGBDFs (Fig.~\ref{fig:mwlP}), the same $\rh=0.3$ and $0.2$~pc models, respectively, are consistent with the observations, being again larger than for G-dwarfs.

\subsubsection{Eccentricity distribution}
\begin{figure*}
 \begin{center}
 $\begin{array}{cc}
   \includegraphics[width=0.45\textwidth]{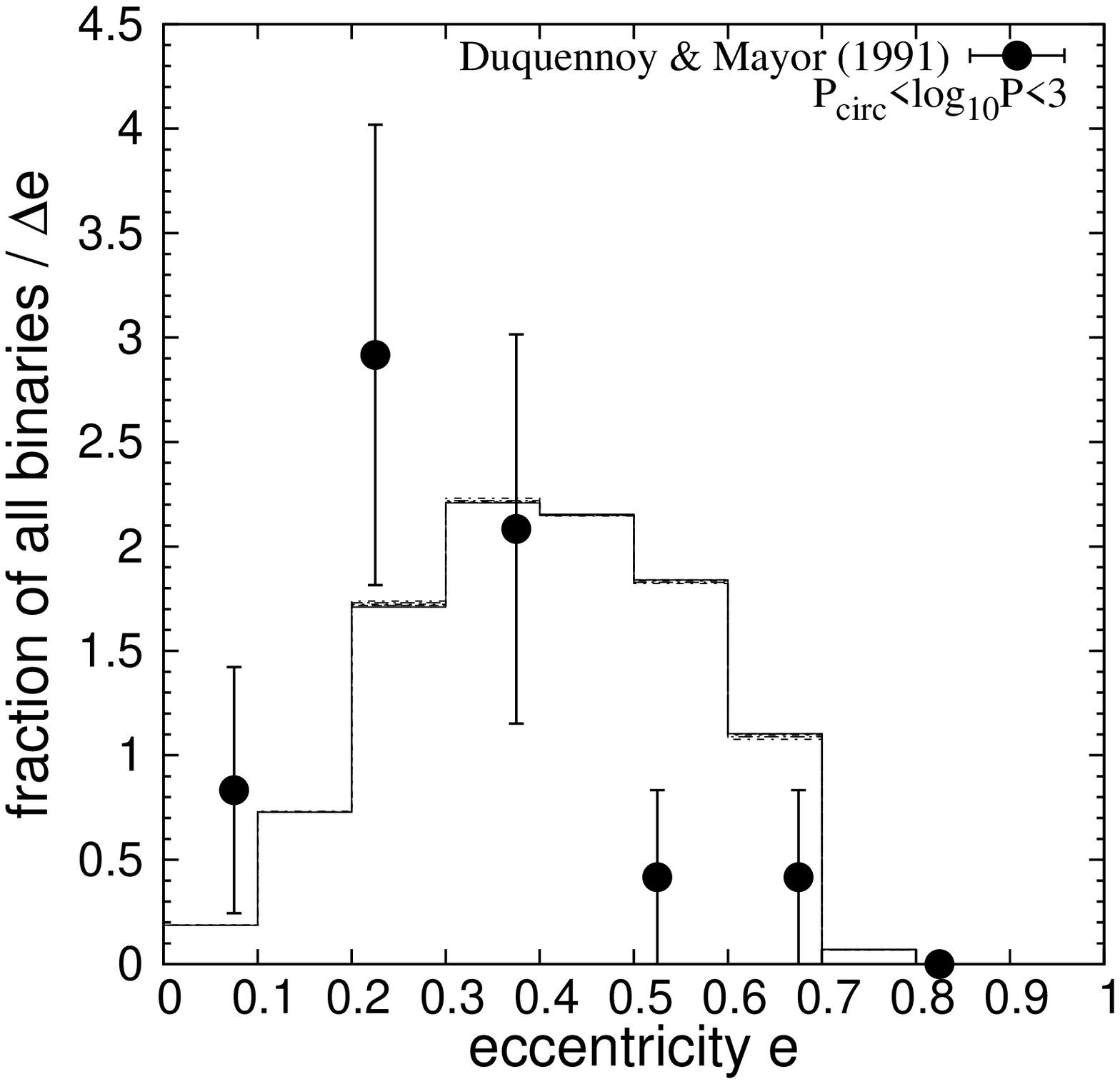} & \includegraphics[width=0.45\textwidth]{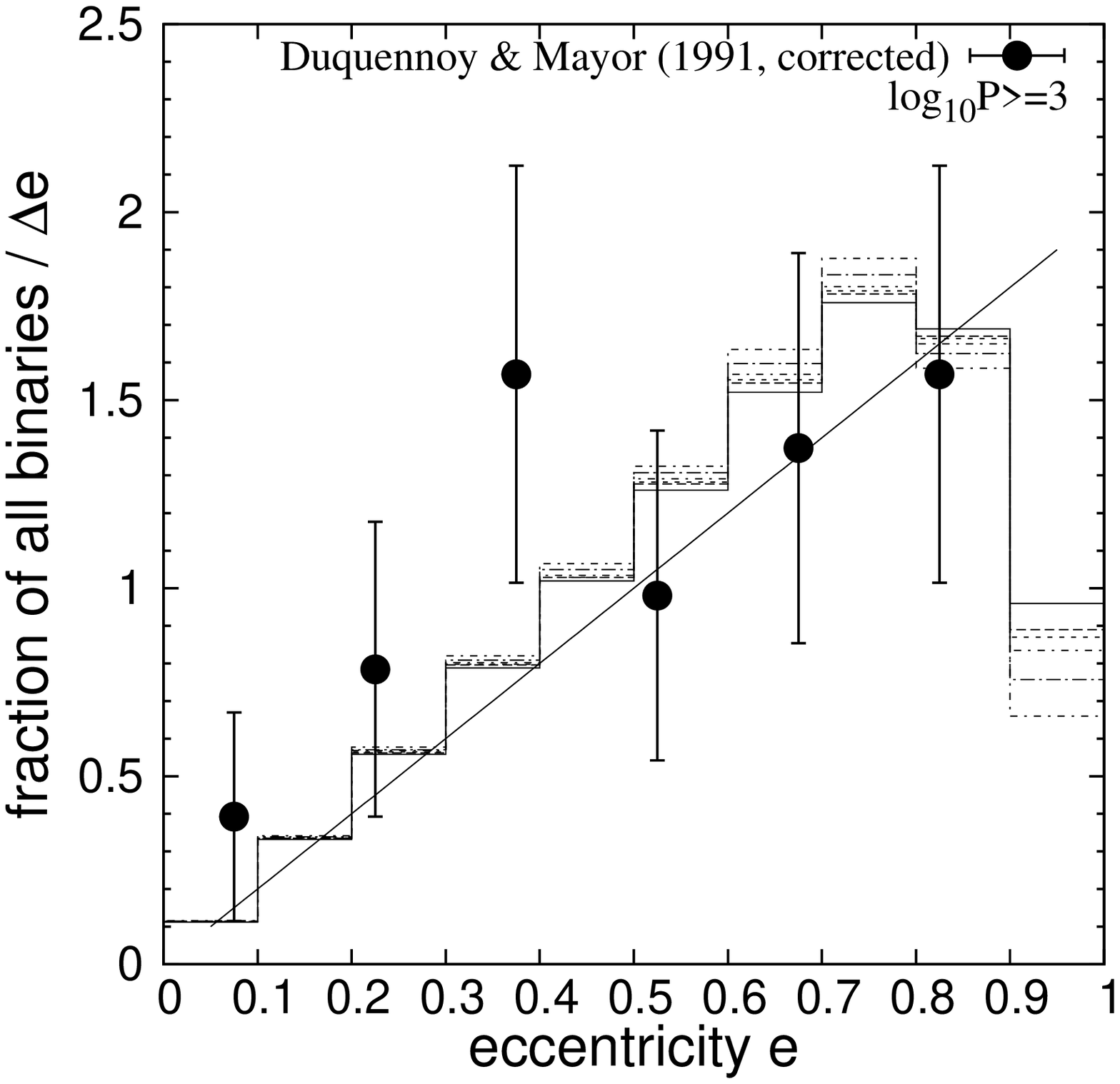}
 \end{array}$
 \end{center}
  \caption{Eccentricity IGBDFs for G-dwarf binaries in comparison with the observations by DM91 in the indicated period ranges. $P_{\rm circ}=11.6$~d is the circularization period derived from the observational data. Note that in both panels the data is (exceptionally) normalized to the total number of binaries instead of systems, in order to be able to compare to the observations. This is also why IGBDF models with different $\rh$ (histograms, different line-types) can hardly be distinguished. \textbf{Left panel:} The eccentricity BDF is bell-shaped for orbital periods below $10^3$~d due to pre-main sequence eigenevolution as in the observational data. \textbf{Right panel:} For $P\geq10^3$~d the $e$-distribution follows the thermal distribution ($f_b(e)=2e$, solid line) for the IGBDF model as well as in the observations. The thermal eccentricity distribution is invariant of stimulated evolution.}
 \label{fig:mwe}
\end{figure*}
The eccentricity IGBDF for G-dwarf binaries is in agreement with the observational data (Fig. \ref{fig:mwe}). The distribution of eccentricities is bell-shaped for $\lP_{\rm circ}=1.06<\lP<3$, where $P_{\rm circ}$ is the circularization period identified by DM91 below which the orbits are circular ($e\approx0$). Circularisation occurs through pre-main sequence eigenevolution \citep[][Paper~I]{Kroupa1995b}. The eccentricity IGBDF follows the thermal distribution for $\lP>3$ because it is invariant to stimulated evolution.

\subsubsection{Single and binary population in dependence of spectral type}
\label{sec:fbin}
\begin{figure}
 \begin{center}
    \includegraphics[width=0.45\textwidth]{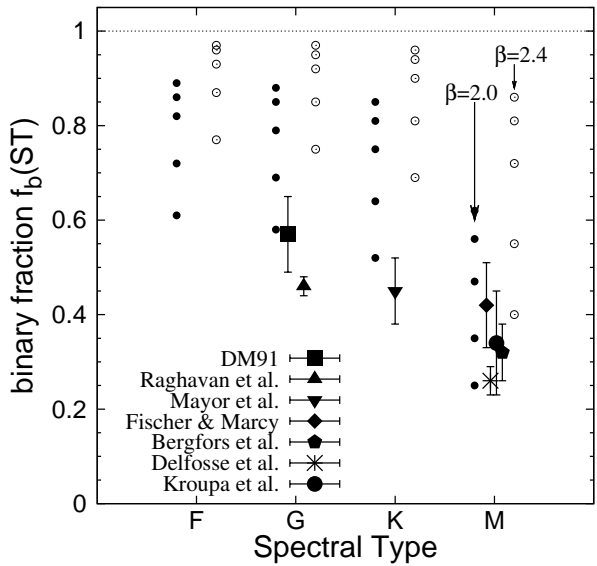}
 \end{center}
  \caption{Comparison of the binary fractions among systems of one spectral type (SpT) for IGBDF models with different $\rh$ (filled and open circles for an ECMF slope $\beta=2$ and $2.4$, respectively, in each column from top to bottom: $0.8,0.6,0.4,0.2$ and $0.1$ pc). The initial binary fraction for all spectral types is $\fb=1$ (dotted horizontal line). M-dwarf binaries have the lowest binary-fraction when they enter the Galactic field. The flatter the ECMF (smaller $\beta$), the lower is the binary-fraction (Sec. \ref{sec:paramstudy}). Observed binary fractions agree with the $\beta=2$, $\rh=0.1-0.2$~pc IGBDF model best \citep[data taken from DM91;][]{Mayor1992,FischerMarcy1992,Kroupa1993,Delfosse2004,Raghavan2010,Bergfors2010}.}
 \label{fig:mw_fb_st}
\end{figure}

The fraction of binaries in the field is a function of the spectral type (i.e. mass) of the primary. The later the primary spectral type the lower is the resulting binary fraction in the field (Fig.~\ref{fig:mw_fb_st}). Especially M-dwarfs have a lower binary-fraction than binaries with a F-, G- or K-type primary. The reason for this is twofold.

First, due to the shape of the stellar IMF M-type stars are most numerous so that upon random pairing of the binary components at birth and after eigenevolution (Sec.~\ref{sec:model}) about $\approx90$~per~cent of all initial binaries carry a companion of spectral type M. Thus, everytime a binary dissolves, in 9 out of 10 cases at least one M-dwarf will end as a single star (two if a M-dwarf binary dissolves). Each M-dwarf contributes to $\ncms({\rm M})$, and therefore $\fb({\rm M})=\nb({\rm M})/\ncms({\rm M})$ shrinks.

Secondly, binaries with a lower binding-energy, $\Eb$, are more prone to dissolution (Paper~I) and the binding-energy is proportional to the mass of the primary, $\Eb\propto m_1$. Since, by construction, all binaries intially follow the same initial period distribution \citep[Fig.~\ref{fig:mwlP}]{Kroupa1995b}, the M-dwarf (low $m_1$) energy IGBDF is shifted to slightly lower energies compared to F-, G- and K-type binaries. Thus, it is generally easier to dissolve M-dwarf binaries than binaries with a primary of an earlier spectral type. Furthermore, low-mass binaries are more frequent, again owing to the shape of the stellar IMF. In the model, $\approx57$~per~cent of all binaries have a M-dwarf primary initially. Dynamical encounters including M-dwarfs will therefore occur often in the clusters. In turn, each disruption of a M-dwarf binary will reduce $\nb({\rm M})$ by one, increase $\ncms({\rm M})$ by one and, in turn, reduce $\fb({\rm M})$.

All observational data in Fig.~\ref{fig:mw_fb_st} again compare best with star formation in compact star clusters $(\rh=0.1-0.2$~pc and $\beta=2)$. Although all stars are locked up in binaries initially, more than half of all systems in the Galactic field end up as single stars due to stimulated evolution in star clusters before they dissolve (Fig.~\ref{fig:paramstudy}). M-dwarfs in the IGBDF model constitute about $\approx80$~per~cent of the single star population in the Galactic field. Only $\approx13$, $3$ and $2$~per~cent of all single stars in the IGBDF model have spectral type K, G and F, respectively (the remaining $2$~per~cent are of spectral type A, for a Galactic field population which consists of systems with $m_1\leq2\;\msun$, Sec.~\ref{sec:extraction}).

For $\rh\approx0.1-0.2$~pc and $\beta=2$ the single star fraction, $\fsing({\rm SpT})=1-\fb({\rm SpT})$, in the model becomes $\fsing({\rm M})\approx0.75-0.85$, $\fsing({\rm K})\approx0.36-0.48$, $\fsing({\rm G})\approx0.31-0.42$ and $\fsing({\rm F})\approx0.28-0.39$. The total single star fraction amounts to $55-66$~per~cent and is in excellent agreement with the estimate by \citet{Lada2006}, that about $2/3$ of all primary stars are single.

\section{Discussion \& model predictions}
\label{sec:predict}
\subsection{Formation in compact star clusters}
\label{sec:compact}
Comparison of the observational data with the model has suggested that MW star clusters typically formed quite compact ($\rh\approx0.1-0.3$~pc, Sec.~\ref{sec:MW}). Such small radii compare with the observational lower end of the sizes of dense cores in giant molecular clouds. However, the sizes of these dense cores range up to $2$~pc and the spatial extends of embedded clusters are typically comparable \citep[e.g.][]{Lada2003}. But there is evidence that the forming stars within the embedded cluster start dynamically cold \citep{Walsh2004,Peretto2006,Lada2008} and that protostellar objects are more confined than more evolved young stellar objects \citep{Teixeira2006,Muench2007}. Thus, an embedded cluster will collapse to a smaller configuration. In that sense the here derived half-mass radii may be interpreted as the sizes of clusters when they reach their peak stellar density and stimulated evolution is most efficient.

Additionaly the IGBDF model implicitely assumed that a typical half-mass radius for all clusters exists (Sec.~\ref{sec:igbdf}). If a mass-radius relation for embedded clusters better describes reality \citep[e.g.][for virialised gas cores in giant molecular clouds]{Harris1994}, the typical $\rh$ for MW clusters can be seen as an average value for all clusters of any mass and size. Note that if such a mass-radius relation exists, upon averaging the inferred best $\rh$ will be closer to the true value for low-mass clusters, which are most important for the Galactic field binary population (Sec.~\ref{sec:importance}). Higher mass clusters will then occupy a somewhat larger radius range. However, a trend of radius with luminosity (or mass) in young star clusters has been shown to be shallow for the galaxy merger NGC~3256 \citep{Zepf1999}, for stellar clusters in 18 spiral galaxies \citep{Larsen2004} and for young clusters in M 51 \citep{Scheepmaker2007}.

The typical $\rh$ is arrived at assuming $\beta=2$ down to $5\msun$. The ECMF might however flatten ($\beta\rightarrow0$) below $\approx50-100\msun$ \citep{Lada2003}, implying that fewer low-mass clusters are present compared to the numbers used in the model. According to a computation with a broken power-law ECMF, where $\beta=0$ for $\mecl\leq100\msun$ and $\beta=2$ otherwise was chosen, typical half-mass radii nedd to be larger by $\approx0.1$~pc in order to agree with the observations.

A small difference in the solutions for $\rh$ ($0.1$ vs. $0.3$~pc) in the solutions for G- and M-dwarf binaries might be evident (Fig.~\ref{fig:mwlP}). Since the estimated cluster size is also a measure of how dense the region is in which the respective sub-population has formed, the possible difference might simply indicate their formation in different locations of the same cluster. A primordial mass-segregated cluster, where the G-dwarfs would be more centrally confined to a region of higher density while M-dwarfs form out to larger radii, would naturally account for the apparently different ranges of allowed cluster radii.

The uncertainty in the inferred typical cluster size of $0.2$~pc does not appear to be very large, given the estimates for $\rh$ by comparison of the model with independent observations (Sec.~\ref{sec:MW}). Even if a possible error in the ECMF index, $\beta$, of up to $0.5$ is considered \citep{Larsen2009} the uncertainty in $\rh$ is of order $0.1-0.3$~pc only (Fig.~\ref{fig:paramstudy}, left panel).

To calculate the composition of the Galactic field stellar population a (constant, average) global SFR was adopted. A declining SFR history might instead be better suitable to describe the evolution of the MW disc \citep{BoissierPrantzos1999,NaabOstriker2006,SchoenrichBinney2009}. If early on the SFR has been higher than average, clusters more massive than allowed for the adopted SFR (eq.~\ref{eq:weidner}) would have been able to form, which would have contributed a larger number of single stars. Later, when the SFR sank below the average SFR the highest mass clusters wouldn't be able to form any more and more binaries would enter the field originating in the low-mass clusters. These effects might eventually compensate each other, but will depend on the actual history, which can't be tested without a modification of the used code. Also, the global history might be non-representative for the solar neighbourhood \citep{BoissierPrantzos1999}. However, even if the SFR has been much higher in the past only and settled to the present value, the inferred typical $\rh$ wouldn't change strongly. Reducing or enhancing the SFR by up to two orders of magnitude and at the same time retaining the observed binary-fraction requires, respectively, a $\rh$ smaller by $\approx0.1$~pc or larger by $\approx0.2-0.3$~pc only (as evident from Fig.~\ref{fig:paramstudy}, middle panel).

In this sense, the typical cluster size seems to be well constrained by the IGBDF model.

\subsection{MW orbital-parameter BDFs}
\label{sec:mwbdf}
\begin{figure*}
 \begin{center}
 $\begin{array}{cc}
   \includegraphics[width=0.45\textwidth]{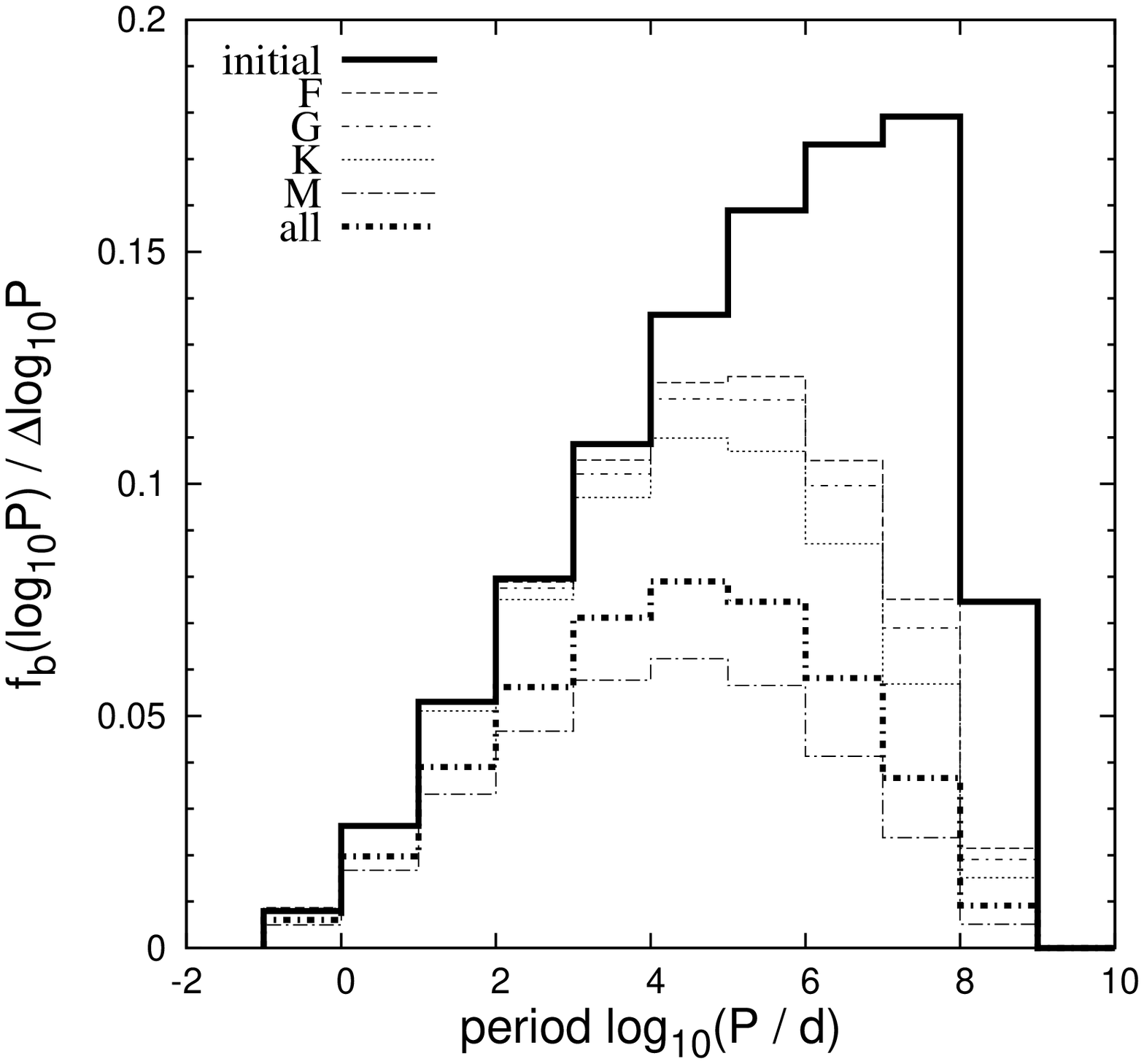} & \includegraphics[width=0.45\textwidth]{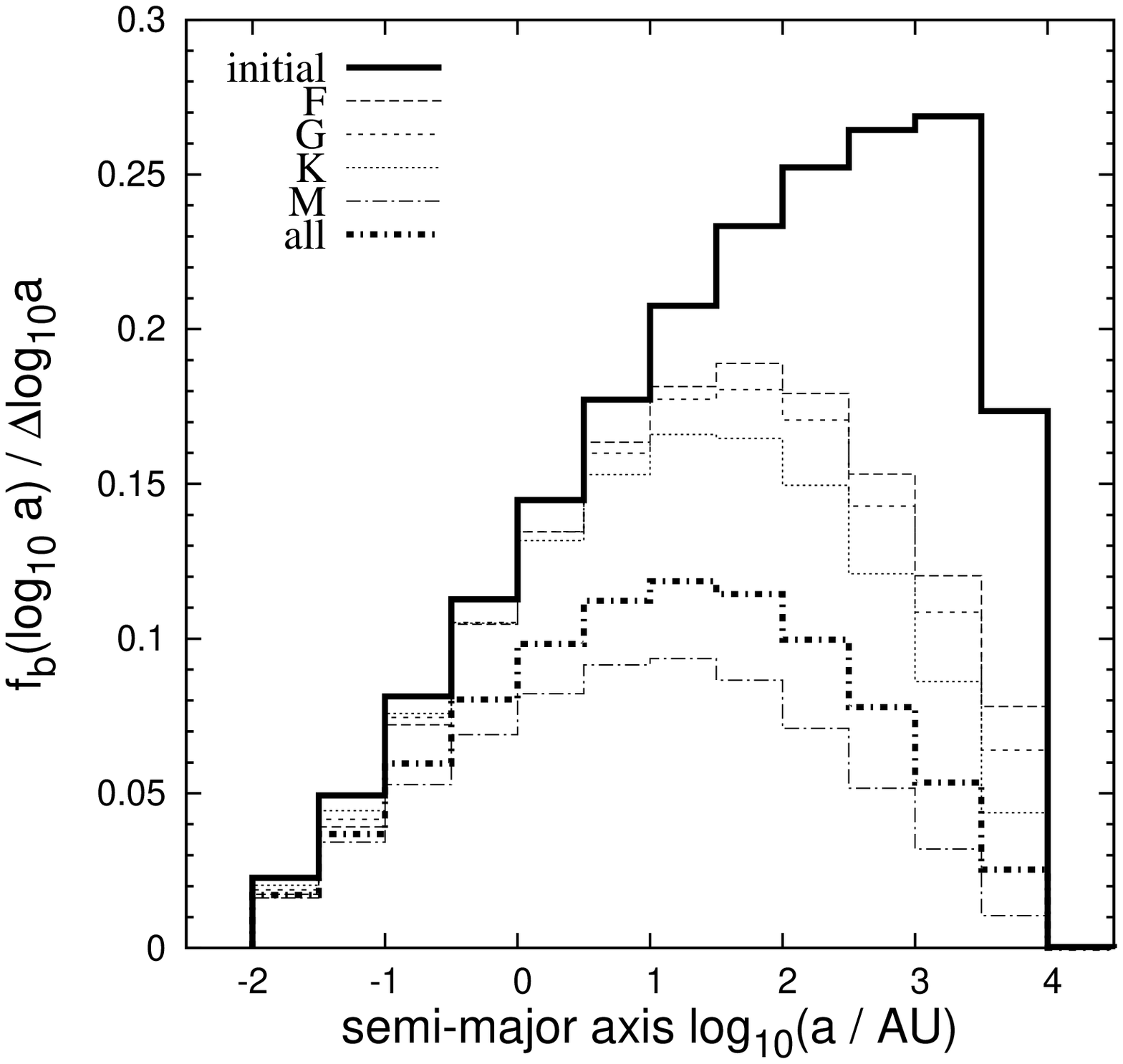} \\
   \includegraphics[width=0.45\textwidth]{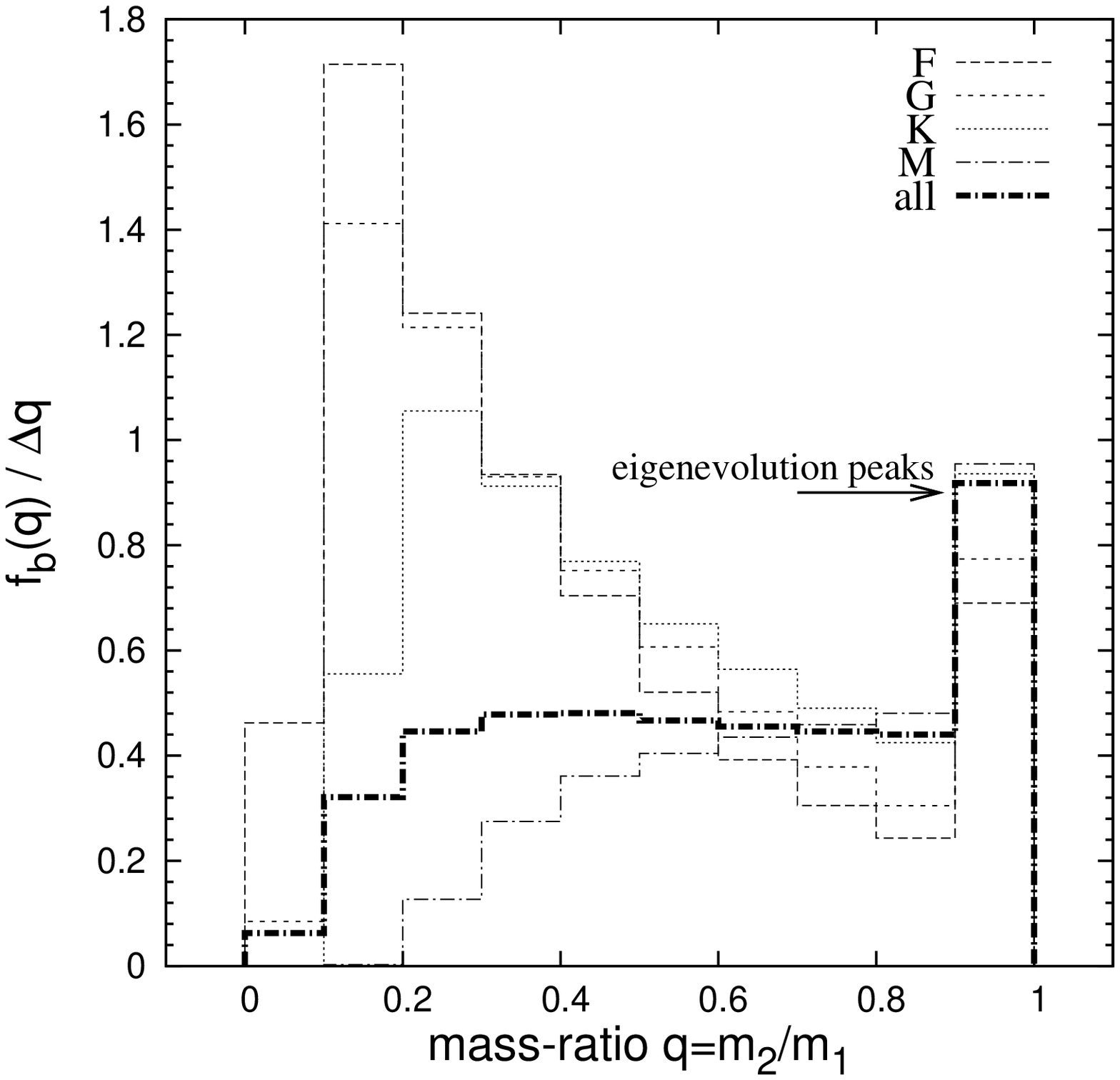} & \includegraphics[width=0.45\textwidth]{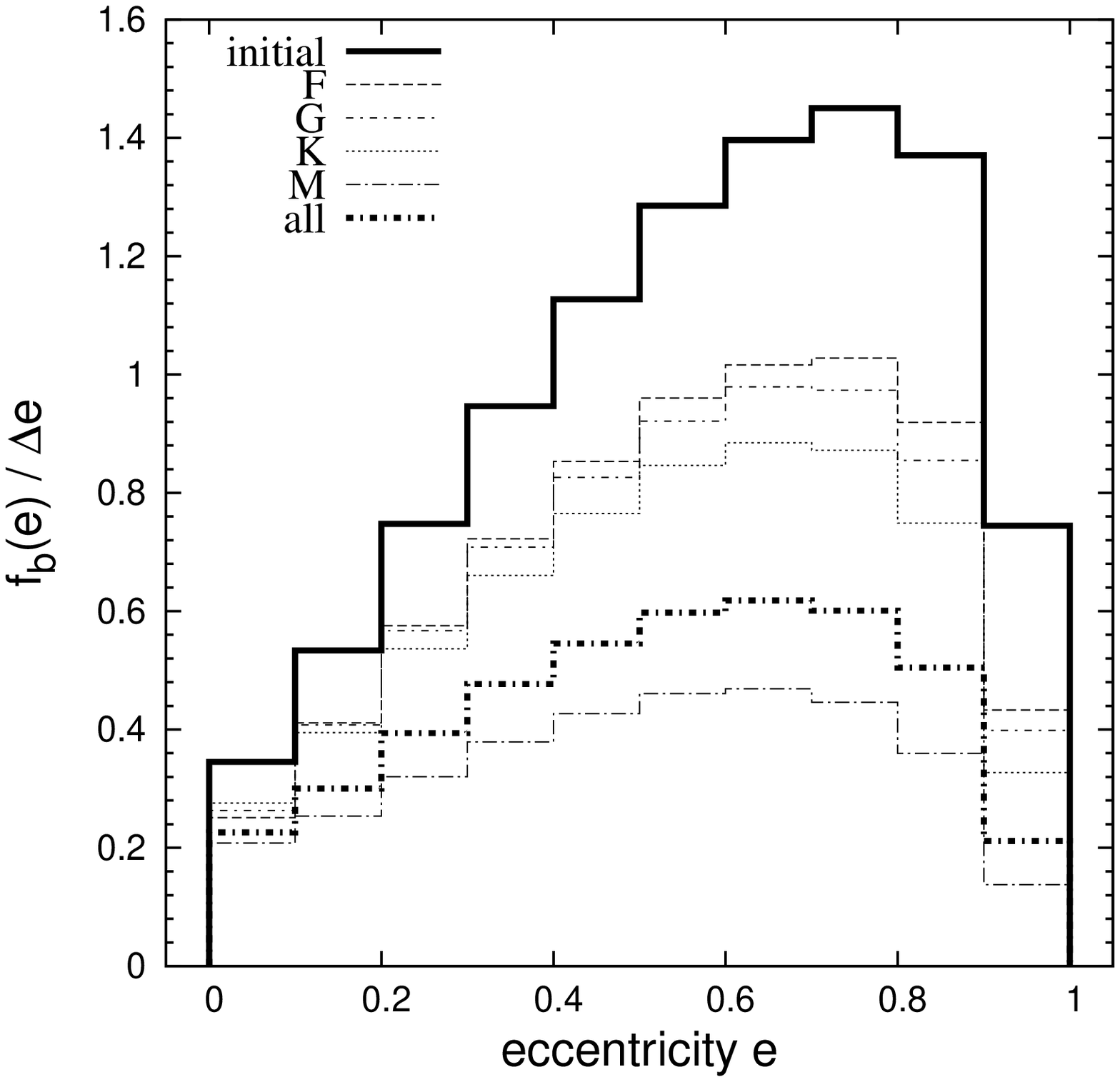}
 \end{array}$
 \end{center}
  \caption{Predictions for the period ($\lP$), semi-major-axis ($a$), mass-ratio ($q$) and eccentricity ($e$) IGBDFs for different spectral types and the combined distributions in the solar neighbourhood for the MW models (Sec. \ref{sec:MW}) which best fit the observational data ($\rh\approx0.2$~pc). The $\lP$-, $a$- and $e$-IGBDFs for F-, G- and K-binaries appear very similar, but can probably be distinguished at low and high $q$-values in the mass-ratio IGBDF. The peaks seen for $q=0.9-1$ in the mass ratio distributions are due to pre-main sequence eigenevolution (Sec.~\ref{sec:model}). All IGBDFs for M-dwarfs and the all-primary mass IGBDFs are distinct from those for F- to K-binaries. Note that a common initial distribution for the $q$-IGBDFs can not be given since they are different for different spectral types (see discussion in Sec. \ref{sec:predict}).}
 \label{fig:predict}
\end{figure*}
Since the IGBDF models for the MW (Sec. \ref{sec:MW}) with typical star cluster sizes of $\rh\approx0.2$~pc agree well with \emph{independent} observational data, these models are used to predict the period, semi-major axis, mass-ratio and eccentricity IGBDFs for the solar neighbourhood.

Fig. \ref{fig:predict} depicts the resulting distributions for binaries of different spectral type and the combined distributions ($m_1\leq2\msun$; no additional cuts, e.g. in period, are applied). For the period, mass-ratio and eccentricity IGBDFs the distributions for F-, G-, and K-type binaries are very similar and therefore probably hard to distinguish by observations. However, the mass-ratio IGBDF might hold the ability to test the model prediction for different spectral types at low ($q\approx0.1-0.2$) and high ($q\approx0.9-1$) mass-ratios since differences are more pronounced there. In particular, there is a larger gap between the G- and K-binaries around $q\approx0.15$ and between the F- and later type binaries at $q\approx0.05$ which might be visible in observations, too.

The M-dwarf binary IGBDFs for all quantities are distinct from the corresponding distributions for earlier spectral types since breaking-up of binaries having at least one M-type component happens frequently before their birth clusters dissolve (Sec.~\ref{sec:fbin}). Thus, by the time the M-dwarfs emerge from the clusters their majority are single stars consistent with observations \citep[Sec. \ref{sec:fbin}]{Lada2006}, despite being born as binaries. Differences in observed distributions for M-dwarf binary populations and those of later types should be apparent and are thus suited to test the IGBDF predictions. The combined IGBDFs lie typically between the M-dwarf distribution and the earlier types.

Note that it is not possible to show a common initial mass ratio distribution since it depends, in contrast to the other IGBDFs, on the considered spectral type \citep[see also][]{Kouwenhoven2009}. This is a result of the random selection of \emph{birth} binary component masses and the upper limit for the mass of the primary star when a spectral type-limited sample is investigated\footnote{By definition the secondary mass can only be lower.}. While at least the shape of the initial mass ratio distributions is similar for F-, G- and K-binaries it is completeley different for M-dwarfs (compare, e.g., the middle with the right panel in Fig.~\ref{fig:mwmr} for the initial distributions of G- and M-dwarfs, respectively). The IGBDFs \emph{decline with increasing mass ratio for F-, G- and K-type binaries}, while \emph{the mass ratio IGBDF for M-dwarfs is increasing with increasing~$q$}. We explicitly note that the same observational data for M-dwarfs, for which typically a flat $q$-distribution is inferred, is consistent with the increasing trend in the models. The mass ratio distribution is \emph{flat only if all primary-masses are combined to construct a mass ratio IGBDF}.

As opposed to DM91, \citet{Raghavan2010} find a mass ratio distribution for binaries with a solar-type primary that is flat between $q\approx0.2$ and $0.9$, while finding a similar period distribution as DM91. This can not be expected within the framework of the IGBDF model. Since the range of considered primary masses in \citet{Raghavan2010}'s and DM91's study are comparable, their mass ratio distributions should be in agreement if the period distributions are, and vice versa. A flat distribution from the models is only obtained if the complete late-type binary population is considered (Fig. \ref{fig:mwmr}, left panel, and Fig.~\ref{fig:predict}). Although this apparent inconsistency might be due to model assumptions (see Sec.~\ref{sec:limitation}), the presented results are otherwise very successfull in describing independent observational data for the MW, adjusting only one free parameter (the typical half-mass radius $\rh$, Sec. \ref{sec:MW}). It is noted that DM91 monitored radial velocities for their sample over a period of 13 years finding accurate orbital solutions, while \citet{Raghavan2010} compile data from various sources covering many techniques. However, in order to understand these issues better it will be necessary to study the consistency between the \citet{Raghavan2010} and the \citet{Reid1997} data - how can both, the (essentially) G-dwarf and the all-primary combined mass ratio distributions be flat at the same time?

\subsection{Dependence on galaxy morphology}
\label{sec:morph}
\begin{figure*}
 \begin{center}
 $\begin{array}{cc}
   \includegraphics[width=0.45\textwidth]{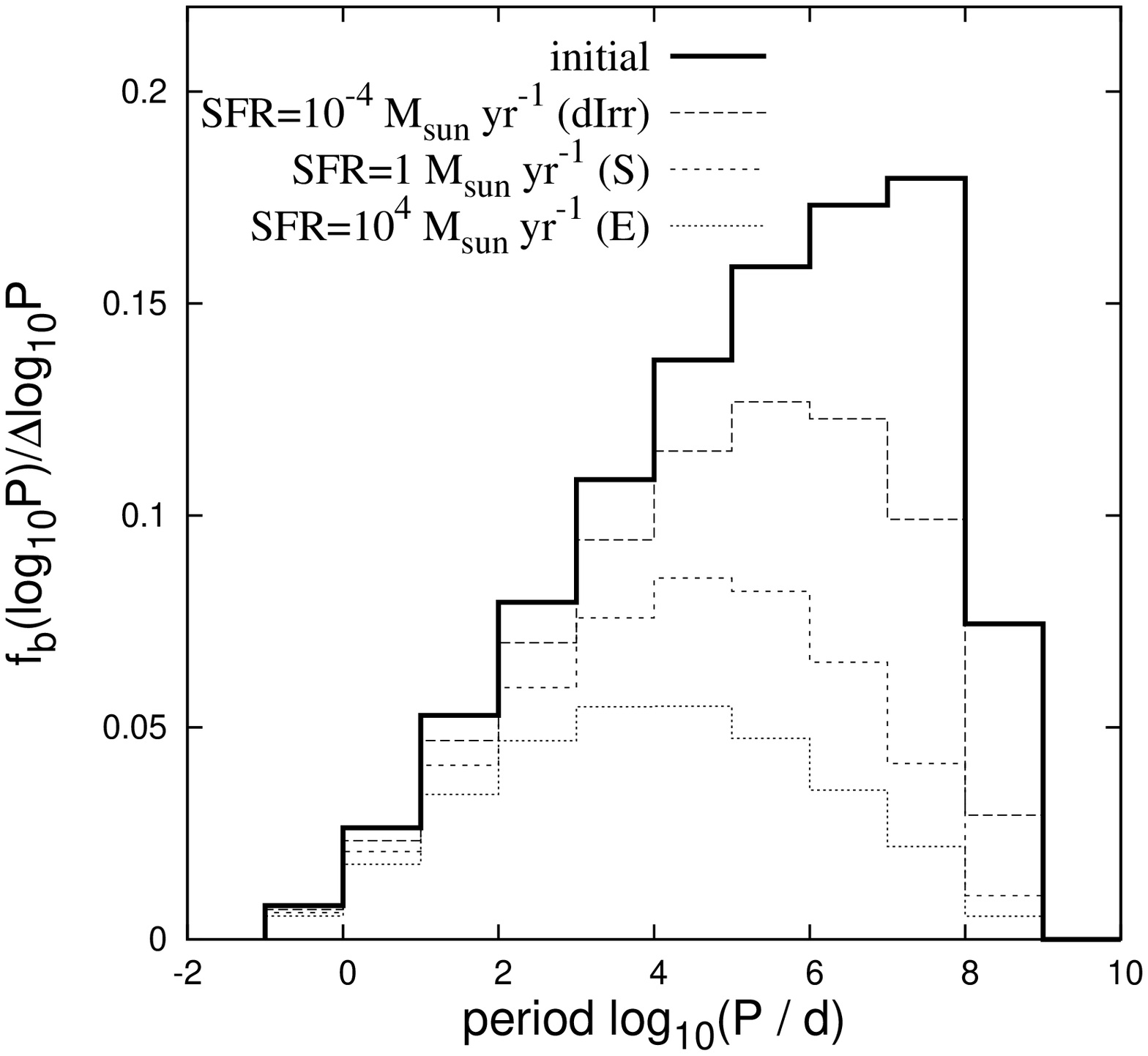} & \includegraphics[width=0.45\textwidth]{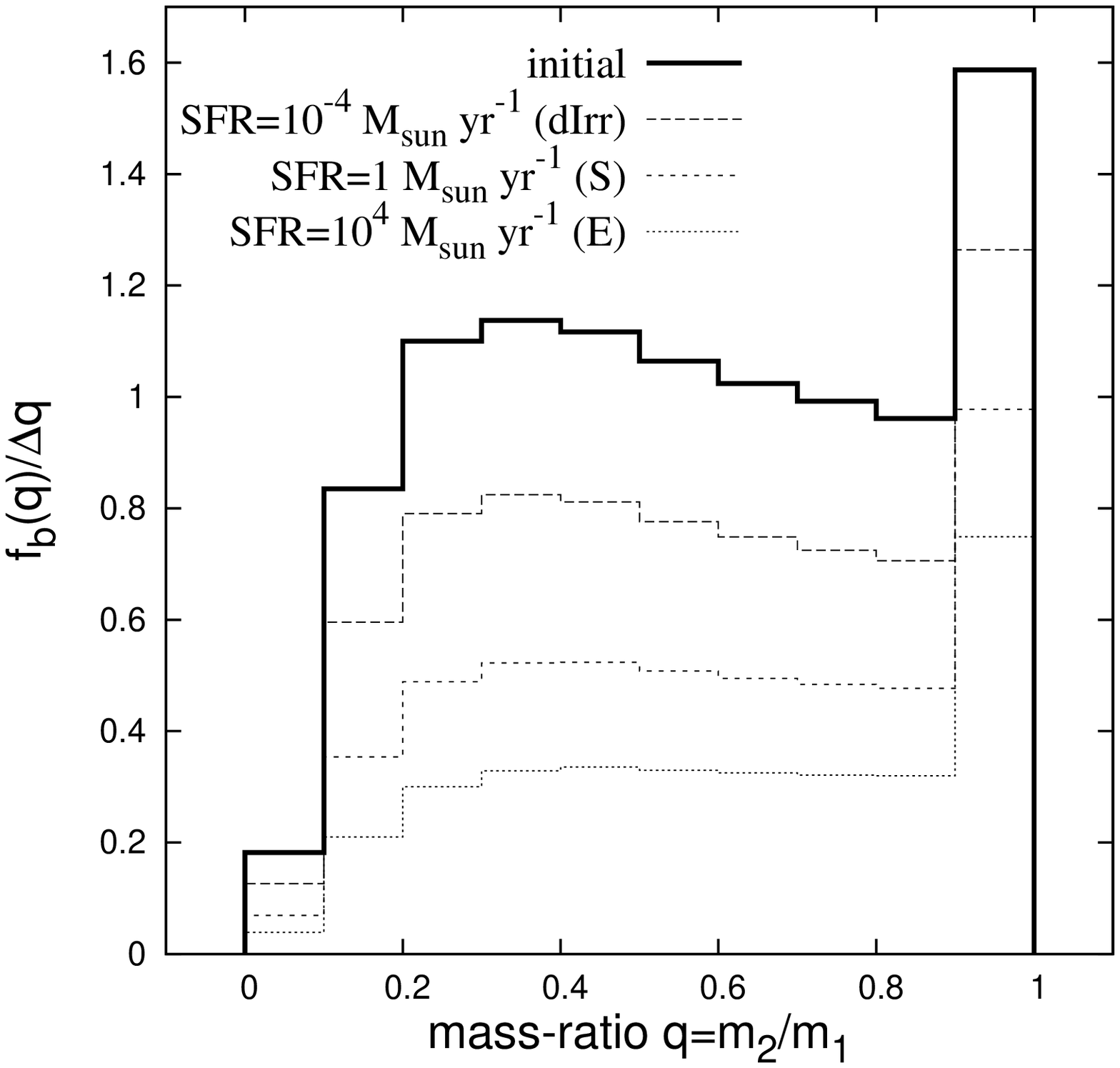}
 \end{array}$
 \end{center}
  \caption{Predictions for the period ($\lP$) and mass-ratio ($q$) IGBDFs in dependence of galaxy morphology. Model parameters are as those for the MW (Sec.~\ref{sec:MW}, $\rh=0.2\;$pc) except that SFRs are used according to the galaxy type (Sec.~\ref{sec:morph}). Elliptical (E) galaxies have the lowest binary-fraction (the area below the distribution) and the period IGBDF peaks at shorter periods than for spirals (S) and dwarf~irregulars (dIrr). The shapes of the mass ratio IGBDFs for the different galaxies resemble each other. The pre-main sequence eigenevolution peak in the mass ratio IGBDF is visible, as in Fig.~\ref{fig:predict}.}
 \label{fig:morph}
\end{figure*}
The binary properties of galaxies within the IGBDF model are strongly dependent on the SFR (Sec.~\ref{sec:paramstudy}). Since SFRs in galaxies are observed to cover a large range from $\approx0\;\mpyr$ for giant ellipticals to $\gtrsim1000\;\mpyr$ for ultra luminous infrared galaxies \citep[ULIRGS,][]{Grebel2011}, binary-frequencies and binary-properties are expected to vary between galaxies of different morphology. This might have cosmological implications, such as for the SN~type~Ia rates in galaxies.

\subsubsection{Elliptical galaxies}
Ellipticals (Es) are nowadays more or less free of cold gas and are pressure- or random-stellar-motion-supported with low or no star formation activity. Even the suspected precursors of giant ellipticals, quasars at high redshift ($z\approx6$, $t_{\rm universe}\lesssim1$~Gyr), reveal supersolar metallicities indicating a starburst that quickly enriched the material with metals \citep{Fan2001}. This suggests that Es had a large SFR initially until their gas reservoir was depleted. One of the highest-redshift quasars known has a SFR of $\approx1000\mpyr$ as derived for ULIRGS \citep{Fan2006}. For SFR$=10^3\;\mpyr$, from the middle panel of Fig.~\ref{fig:paramstudy}, a binary fraction of the order $\approx30-40$~per~cent for a typical cluster size of $\rh=0.2$~pc can be inferred. It is noted that a non-shallow mass-radius relation (Sec.~\ref{sec:compact}), which is not considered in the present models, might affect the results for such high SFR. If, additionally, during star bursts low-mass clusters are not able to form, i.e. $\meclmin$ is larger, this would further lower the binary-fraction in Es (Fig.~\ref{fig:paramstudy}, right panel). \emph{Thus, if E galaxies formed rapidly they ought to have low binary fractions}.

\subsubsection{Spiral galaxies}
In terms of the SFR, spiral galaxies are intermediate objects between Es and dwarf galaxies. SFRs in spirals like the MW lie between $0.1$ and $10\mpyr$ \citep[e.g.][]{Lee2009,Lee2011}. For the same cluster size and a SFR of $1\mpyr$, from Fig.~\ref{fig:paramstudy} a global binary frequency of $\approx40-50$~per~cent is expected, similar to what is seen in the solar neighbourhood (Sec.~\ref{sec:MW}).

\subsubsection{Dwarf irregular galaxies}
Dwarf irregular (dIrr) galaxies have very low SFRs ranging down to $10^{-5}\;\mpyr$ \citep{Lee2009,Lee2011}. For $\rh\approx0.2$~pc dIrrs would be expected to exhibit a significantly larger binary fraction, which is of the order of $70-80$~per~cent.

\subsubsection{IGBDFs for galaxy types}
The above results hold if cluster formation in the MW is representative for other galaxy types (Sec.~\ref{sec:MW}, $\rh=0.2\;$pc, but with different SFRs). Then, the expected period and mass ratio IGBDF for ellipticals, spirals and dIrrs for a population of binaries with $m_1\leq2\msun$ (no cuts in period or primary-mass) are expected to be as in Fig.~\ref{fig:morph}. The period IGBDF of Es peaks at shorter periods than the corresponding distributions for S and dIrr galaxies. It is due to the initial star burst in Es (high SFR), namely that more dynamically evolved binary populations from high-mass clusters, which are not present in the lower SFR spirals and dIrrs (eq.~\ref{eq:weidner}), contribute to the galactic field of Es (Sec.~\ref{sec:paramstudy}). The mass ratio IGBDFs for all morphological types look alike, a difference only appearing through the varying binary-fraction between the galaxies (the area below the distributions).

\subsection{Model Limitations}
\label{sec:limitation}
The validity of the results and predictions outlined in this analysis depend on the accuracy of the assumptions entering the IGBDF model. While the model formulated in Sec.~\ref{sec:igbdf} appears robust, the results obtained by performing the integration in eq.~(\ref{eq:igbdf}) depend on the properties of the \emph{birth} and \emph{initial} binary population \citep[see Sec.~\ref{sec:model},][Paper~I]{Kroupa1995a,Kroupa1995b} and the resulting analytical description of the evolution of binary properties in the $N$-body models of Paper I, which use this initial binary population. The two major assumptions are that (i) the birth binary population is formed via \emph{random-pairing} of the two binary components from the canonical stellar IMF and that (ii) the birth population evolves into the initial population via \emph{pre-main sequence eigenevolution}, which was specifically parameterised for G-dwarf binaries \citep{Kroupa1995b}. If, e.g., instead of the declining mass ratio distribution found in DM91 the flat mass ratio distribution in \citet{Raghavan2010} for solar-type stars were correct, this would possibly imply a birth pairing method for binaries which is different from random-pairing and, in turn, would require alteration of the binary birth population \citep[see also][]{Kouwenhoven2009}. Alternatively, the eigenevolution model might need adjustments, e.g. through primary star dependent eigenevolution parameters $\lambda=\lambda(m_1)$ and $\chi=\chi(m_1)$, so that the initial binary population may be mildly different for other spectral types. These limitations should always be kept in mind, but the overall results of the IGBDF model would not be affected.

\section{Summary \& Outlook}
\label{sec:sum}
Following observational evidence which implies all stars less massive than about $2\msun$ to form as binaries with component masses picked randomly from the stellar IMF in discrete star formation events (i.e. embedded star clusters) and allowing for pre-main sequence eigenevolution (Sec.~\ref{sec:model}) and stimulated evolution, the concept of \emph{integrated galactic-field binary distribution functions}~(IGBDFs) is introduced. Adding up the stellar populations that ever formed in star clusters which are selected from an ECMF yields the statistical binary properties of a galactic field population (Dynamical Population Synthesis). This approach is similar to the IGIMF theory which adds up the stellar IMFs in individual clusters to calculate the galaxy-wide IMF (Sec.~\ref{sec:intro}).

The IGBDFs depend on the minimum cluster mass, $\meclmin$, the galaxy-wide SFR, the steepness of the ECMF and the typical, or average value for cluster half-mass radii, $\rh$, in a star cluster system. The galactic field binary-fraction increases with decreasing $\meclmin$ and SFR, with increasing index $\beta$ of the ECMF (eq.~\ref{eq:ecmf}) and with increasing $\rh$ (Fig.~\ref{fig:paramstudy}). It is found that low-mass (i.e. low-density) clusters contribute most binaries to the field since stimulated evolution is least effective in them and low-mass clusters are most numerous due to the steep ECMF. High-mass (high-density) clusters donate most single stars since binary destruction is effective and the number of stars forming in a high-mass cluster is larger than in a low-mass cluster.

Applying the IGBDF model to the MW, i.e. estimating $\meclmin,\;\beta$ and the SFR from observations, the period, mass-ratio and eccentricity IGBDFs are constructed for different typical cluster sizes, $\rh$. The models independently agree with observed distributions for late-type binaries in the solar neighbourhood, solely adjusting the typical $\rh$ to $0.1-0.3$~pc, which is the single remaining free model parameter. This suggests that MW clusters typically form quite compact. M-dwarf binaries appear to have formed in slightly more extended clusters than G-dwarfs, a possible sign of mass-segregation. The integrated populations show that the majority of all Galactic field primaries end up being single stars despite being born in binary systems. In particular, the Galactic field binary fraction is a function of the spectral type, i.e. $\fb({\rm M})<\fb({\rm K})<\fb({\rm G})<\fb({\rm F})$, and the binary-frequencies derived from binary-star formation and stimulated evolution in compact clusters agree with observational data.

It has been pointed out that the shape of the mass ratio distribution depends on the considered primary-mass (Fig.~\ref{fig:mwmr}). While F- to K-type binaries show a decreasing trend with increasing~$q$, according to the model the M-dwarf mass ratio distribution increases as $q$~increases. A flat distribution is only obtained if primaries from the full mass-range are combined when constructing the mass ratio distribution.

Using the best-fitting model, the named integrated distributions for late-type binary populations (F to M) in the solar neighbourhood are predicted (Fig.~\ref{sec:predict}). While the IGBDFs for F-, G- and K-type binaries are probably hard to distinguish by observation, M-dwarf binary- and the all-binary cumulative distributions appear rather distinct from them. Therefore the M-dwarf and cumulative distributions are probably the best populations, in comparison with the ones for F-, G- and K-binaries, to test the IGBDF predictions.

Assuming star formation in the MW (in terms of the ECMF and cluster radii) is typical also for other galaxy morphologies, the binary population in elliptical galaxies (high~SFR) is predicted to be significantly smaller than in spiral (intermediate~SFR) and dwarf galaxies (low~SFR) and their period- and mass-ratio IGBDFs are calculated.

The IGBDF model will be extended to allow calculation of binary populations in galaxies which had a strongly varying star formation history, i.e. a time-dependence will be incorporated, via SFR($t$). It will be possible to synthesize binary properties in individual nearby (dwarf) galaxies, and the case of the MW may be revisited. A full synthetic galaxy might be constructed and ''observed'' in the computer to mimic real data in order to provide more sophisticated means of comparison with observations. It will also allow to test their influence on observationally derived properties, especially in populations where binaries cannot be resolved, such as velocity dispersions (used to calculate dynamical masses).
\\\\ \textbf{Acknowledgments}\\
MM was supported for this research through a stipend from the International Max Planck Research School (IMPRS) for Astronomy and Astrophysics at the Universities of Bonn and Cologne. We thank K.~M.~Menten for useful suggestions. We thank D.~Raghavan for discussions and help with their data.

\bibliographystyle{mn2e}
\bibliography{binaries,CMF}
\makeatletter   \renewcommand{\@biblabel}[1]{[#1]}   \makeatother

\label{lastpage}

\end{document}